\documentclass[a4paper,11pt]{article}
\usepackage{jheppub} %
\usepackage{lineno}
\usepackage{xcolor}
\usepackage{soul}
\usepackage{nicefrac} %

\usepackage[stable]{footmisc}
\usepackage[font=small,labelfont=bf,tableposition=top]{caption}
\DeclareCaptionLabelFormat{andtable}{#1~#2  \&  \tablename~\thetable}

\usepackage{mmxref}       %

\arxivnumber{} %

\title{%
Effective string theory on a torus: 
the 3d Ising domain wall
}

\author[1,a,b,c]{David Lima \note{Now at \href{https://inductiva.ai/}{Inductiva}. }}
\author[a,b,c]{J. M. Viana Parente Lopes}
\author[a,b,c,d]{José Matos}
\author[d]{Joao Penedones}
\affiliation[a]{Associate Laboratory LaPMET, 4169-007 Porto, Portugal.}
\affiliation[b]{Departamento de Física e Astronomia, Faculdade de Ciências, Universidade do Porto, rua do Campo Alegre s/n, 4169-007 Porto, Portugal.}
\affiliation[c]{Centro de Física das Universidades do Minho e Porto (CF-UM-PT), Departamento de Física e Astronomia, Faculdade de Ciências, Universidade do Porto, 4169-007 Porto, Portugal.}
\affiliation[d]{Fields and Strings Laboratory, Institute of Physics, École Polytechnique Fédérale de Lausanne (EPFL), Route de la Sorge, CH-1015 Lausanne, Switzerland}

\emailAdd{jose.bouradematos@gmail.com}

\abstract{
We use effective string theory (EST) to describe a toroidal 2d domain wall embedded in a 3d torus. 
In particular, we compute the free energy of the domain wall in an expansion in inverse powers of the area, up to the second non-universal order that involves the Wilson coefficient $\gamma_3$.

In order to test our predictions, we simulate the 3d Ising model with anti-periodic boundary conditions, using a two-step flat-histogram Monte Carlo method in an ensemble over the boundary coupling $J$
 that delivers high-precision free energy data. 
The predictions from EST reproduce the lattice results with only two adjustable parameters: the string tension, $1/\ell_s^2$, and $\gamma_3$. We find $\gamma_3 /|\gamma_3^{\text{min}}|= -0.82(15)$, which is compatible with previous estimates.

}

\begin{document}
\maketitle

\nocite{baffigoIsingStringNambuGoto2023} %
\section{Introduction}

Effective String Theory (EST) provides the universal long-distance description of confining flux tubes and domain walls in a wide class of gauge and spin systems \cite{nambuStringsMonopolesGauge1974a, luscherSymmetrybreakingAspectsRoughening1981a,polchinskiEffectiveStringTheory1991a, aharonyEffectiveTheoryLong2013, dubovskyEffectiveStringTheory2012, dubovskyEvidenceNewParticle2013,  dubovskyTheoryQCDString2016, billoPartitionFunctionInterfaces2006, dubovskyFluxTubeSpectra2015c, conkeyFourLoopScattering2016,berattoDeformationCompactifiedBoson2020a}. In three dimensions, the most general diffeomorphism-invariant and bulk Poincaré-invariant action for the worldsheet theory is given by\footnote{We do not include the Ricci scalar because its integral is topological, nor $K^2$, which vanishes by the equations of motion \cite{Aharony:2011gb,aharonyEffectiveActionConfining2009,dubovskyEffectiveStringTheory2012}.}
\begin{equation}
    S=-\int d^2\sigma\sqrt{h}\left(\dfrac{1}{\ell_{s}^{2}} + 2\gamma_{3}\ell_{s}^{2}K^4+\mathcal{O}\left(\ell_{s}^{4}\right)\right),
    \label{eq:action}
\end{equation} 
where $h_{ij}=\partial_{l}X^{\mu}\partial_{j}X^{\nu} G_{\mu\nu}$ is the induced metric, $G_{\mu\nu}$ the metric in the embedding space, and $K^4 =  ( K^\mu_{ij} K_\mu^{ij} )^2 $  with $K^\mu_{ij}$ the extrinsic curvature, defined as $K_{ij}^{\mu}=\nabla_{i}\partial_{j}X^{\mu}$.  The normalization of the second term is chosen so that the phase shift in two-to-two scattering of branons (or phonons of the domain wall) is given by $2\delta=\ell_s^2 s/4+\gamma_3  \ell_s^6 s^3 +O(s^5)$  (see eq.\;(11) in \cite{miroFluxTubeSmatrix2019}). 
This derivative expansion is controlled by the string tension $1/\ell_s^2$ and  the leading non-universal coefficient $\gamma_3$ which encodes the first imprint of the underlying microscopic theory.

In section \ref{sec:theory}, we compute the EST  partition function for a Euclidean torus worldsheet. In particular, we compute the free energy
as an expansion in inverse powers of the area up to the first non-universal order, where 
the Wilson coefficient  $\gamma_3$ appears.
We clarify the issue of normalization of the partition function \cite{caselleJarzynskisTheoremLattice2016,billoPartitionFunctionInterfaces2006} and obtain the free energy solely in terms of the string tension and $\gamma_3$. For the rectangular torus depicted in figure \ref{fig:simulation set up}, we find a perturbative expansion
\begin{equation}
     F(\tau) = F_\text{U} (\tau)-\dfrac{\gamma_{3}}{ \mathcal{A}^{3}}\dfrac{2 \pi^{6}}{225 }(\tau-\bar{\tau})^{4}E_{4}\left(\tau\right)E_{4}\left(-\bar\tau\right)
+O(\mathcal{A}^{-4})
\label{Fintro}
\end{equation}
where $\mathcal{A}=L_1L_2/\ell_s^2$ is the area in units of the string tension, $u=L_1/L_2$ fixes the imaginary part of the moduli $\text{Im}\tau=u$ of the torus and $E_4$ is the holomorphic Eisenstein series of weight 4. 
The universal part $F_\text{U}$ is given in
\eqref{eq:free energy}, as an expansion in inverse powers of the dimensionless area $\mathcal{A}$. This  follows from the first term in the action \eqref{eq:action} and  therefore does not depend on the Wilson coefficients $\gamma_3,$ etc. 

Assuming an integrable low-energy sector with two-to-two phase shift $2\delta(s)=\ell_s ^2 s/4+\gamma_3 \ell_s ^6s^3$, we use the Thermodynamic Bethe Ansatz (TBA) to compute—\emph{to first order in $\gamma_3$}—both the finite-volume spectrum and the partition function (sec.\;\ref{subsec:subleading TBA}). We then determine the leading and next-to-leading non-universal contributions. The leading term matches the path-integral result  \eqref{Fintro}.%

In section \ref{sec:simulation set up}, we describe the simulation setup for a domain wall in the 3d Ising model and the improvements made to the flat-histogram method. 
In section \ref{sec:numerics}, we test our predictions against Monte Carlo simulations of the 3d Ising model in the ferromagnetic phase. In this model there are domain walls that are described by EST and the lightest particle has a finite mass (equal to the inverse of the correlation length). 
Our main results are an improved determination of $\gamma_3$, 
\begin{align}
    \gamma_3 = -0.82(15)|\gamma_3^{\text{min}}|=-0.00106(18)
\end{align}
which is closer to the S-matrix bootstrap bound than previously reported \cite{baffigoIsingStringNambuGoto2023}.
Having a properly normalized partition function allows us to fit the numerical data for the free energy rather than the ground-state energy, simplifying the measurement of  $\gamma_3$  by eliminating the numerical extrapolation to long strings.

Finally, our work points to the potential usefulness of flat-histogram methods - once suitably modified - for the study of domain walls.
We demonstrate that finite-transverse-volume corrections are larger than EST predicts and, in doing so, uncover the coupling between the bulk massive particle and the domain wall, which we will discuss in detail elsewhere.

\section{Effective strings  }\label{sec:theory}

In this section, we perform analytic computations of some observables using the  EST framework described above.

\begin{figure}[t]
    \centering
    \includegraphics[width=0.66\linewidth]{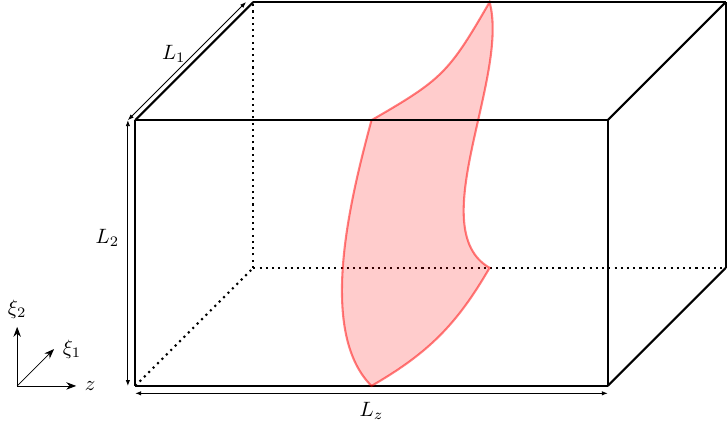}
    \caption{Schematic representation of the domain wall/worldsheet on a three-dimensional torus. 
    For 3d Ising, the boundary conditions are  periodic  along $\xi_1$ and $\xi_2$ and anti-periodic along $z$.    }
    \label{fig:simulation set up}
    \vspace{-0.5cm}
\end{figure}

We focus on the free energy of a finite, non-contractible toroidal worldsheet,
(see fig.\;\ref{fig:simulation set up} for a schematic representation).
We organize the computation in two steps. Firstly, we compute the universal contribution writing the partition function as an infinite sum over string modes with the Goddard–Goldstone–Rebbi–Thorn (GGRT) spectrum \cite{goddardQuantumDynamicsMassless1973}.
Secondly, we compute the leading non-universal contribution (proportional to $\gamma_3$) using 
a path integral approach, which was shown in \cite{billoPartitionFunctionInterfaces2006} to exactly reproduce the universal part.
Previous 
studies of this partition function can be found in 
\cite{caselleJarzynskisTheoremLattice2016,baffigoIsingStringNambuGoto2023,caselleEffectiveStringTheory2024, caristoFineCorrectionsEffective2022}.

Using the low-energy integrability of the $d=3$ worldsheet theory \cite{chenUndressingConfiningFlux2018,Cooper:2014noa,dubovskySolvingSimplestTheory2012}, we present an alternative method for computing the non-universal contribution to the partition function. We first determine the leading $\gamma_3$ correction to the spectrum via the TBA and then sum over the corresponding generalized string modes, in analogy with the GGRT case in subsec.\;\ref{subsec: universal contribution}. At order $\mathcal{O}(\gamma_3/\mathcal{A}^3)$ this reproduces the path-integral result, while at $\mathcal{O}(\gamma_3/\mathcal{A}^4)$ it yields a new prediction that can be tested in Monte Carlo simulations. We expect this method to cease to be valid at $\mathcal{O}(1/\mathcal{A}^5)$, where integrability-breaking effects of order $\gamma_3^2$  set in.

\subsection{Universal contribution: the GGRT spectrum}\label{subsec: universal contribution}

We calculate the universal contribution to the partition function by modeling the system as a one-dimensional object with a center-of-mass degree of freedom in the $z$-direction and string modes as internal degrees of freedom, using the GGRT spectrum. 
These determine the object's rest mass
\begin{equation}\label{eq:GGRT spectrum}
 m_{\text{GGRT}}\equiv
 \sigma L_2
 \mathcal{E}_{k,k^{\prime}}= \sigma L_2\sqrt{1+\frac{4\pi u}{\mathcal{A}}\left(k+k^{\prime}-\frac{1}{12}\right)+\left[\frac{2\pi u\left(k-k^{\prime}\right)}{\mathcal{A}}\right]^{2}},  
\end{equation}
where $k$ and $k^\prime$ are string modes (the left- and right-moving excitations), $\mathcal{A} = \sigma A$ is the area in units of the string tension $\sigma=1/\ell_s^2$, and $u = L_1 / L_2$ is the aspect ratio.
We foliate the worldsheet along $\xi_1$, taking $L_2$ as the size of the corresponding classical string. Foliating in the orthogonal direction would be equivalent to replacing $u$ with $1/u$.

The dispersion relation is
\begin{equation}
    \sigma^2 L_2^2\mathcal{E}_{k,k^{\prime},p_{z}}^2\equiv E^2 = p_z^2 +m_{\text{GGRT}}^2,
\end{equation}
where the prefactor $\sigma L_2$ is the classical string's rest mass. We use $\mathcal{E}$ to refer to the energy density (in string units) $\mathcal{E}\equiv E/(\sigma L_2)$. The full spectrum is then
\begin{equation}\label{eq:qm ggrt spectrum}
 \mathcal{E}_{k,k^{\prime},p_{z}}=\sqrt{1+\frac{4\pi u}{\mathcal{A}}\left(k+k^{\prime}-\frac{1}{12}\right)+\left[\frac{2\pi u\left(k-k^{\prime}\right)}{\mathcal{A}}\right]^{2}+\left(\frac{p_{z}}{\sigma L_{2}}\right)^{2}},
\end{equation}
where it is useful to note that $u / \mathcal{A} = 1 / (\sigma L_2^2)$.
The partition function is the sum over all degrees of freedom, including the string modes $k$ and $k^\prime$, as well as the center-of-mass momentum $p_z$
\footnotemark
\footnotetext{The proper normalization is obtained by starting with a finite-transverse-volume and then taking the infinite volume limit. The transverse momentum is discretized using periodic boundary conditions in the $z$ direction,
\begin{equation}
\exp\left(ip_{z}z\right)=\exp\left(ip_{z}\left(z+L_{z}\right)\right)\Longrightarrow p_{z}=\dfrac{2\pi n}{L_{z}}\qquad\forall n\in\mathbb{Z}. 
\end{equation}

In the limit $L_z \to \infty$, the momentum sum can be approximated by an integral, $\sum_n \to  \frac{L_z}{2\pi} \int dp_z$.}
 \begin{align}
Z_\text{U}  =\text{Tr}\left[e^{-L_{1}H}\right]=&\dfrac{L_{z}}{2\pi}\int dp_z\sum_{k,k^{\prime}}p(k)p(k^\prime)e^{-\sigma L_{1}L_{2}\mathcal{E}_{k,k^{\prime},p_{z}}},\\
=&\left(\dfrac{\sigma\mathcal{A}L_{z}^{2}}{u\pi^{2}}\right)^{1/2}\sum_{k,k^{\prime}}p(k)p(k^{\prime})\mathcal{E}_{k,k^{\prime}}K_{1}\left(\mathcal{A}\mathcal{E}_{k,k^{\prime}}\right).\label{eq:partition function rectangular}
\end{align}
where $p(k)$ is the string level degeneracy, which, for a three-dimensional target space, is given by the number of partitions of the integer $k$.
Integrating over the momentum, we obtain a modified Bessel function and reproduce eq. (2.35) of \cite{berattoDeformationCompactifiedBoson2020a} and eq. (2.23) of \cite{billoPartitionFunctionInterfaces2006}, up to normalization factors.
The modular parameter of the worldsheet torus is $\tau = iu$. To extend the previous computation to general modular parameter $\tau =\alpha+ iu$, one needs to twist the partition function by inserting $e^{-2\pi \alpha L_2 P_2}$ into the trace in eq.\;\eqref{eq:partition function rectangular}.

The large area expansion of the partition function can be derived using the approach outlined in app.\; A of \cite{billoPartitionFunctionInterfaces2006}. 
By expanding both $\mathcal{E}_{k,k^\prime}$ and $K_1$ at large area we obtain 
\begin{equation}
    Z_\text{U} = e^{-\mathcal{A}} \left(\dfrac{\sigma L_z^2}{2\pi u} \right)^{\frac{1}{2}}\sum_{k,k^\prime}p(k)p(k^\prime)\left(1+\sum_{n=1}^\infty \dfrac{f_n(k,k^\prime)}{\mathcal{A}^n}\right)e^{2\pi i \tau \left(k-\frac{1}{24}\right)}
    e^{-2\pi i \bar{\tau} \left(k'-\frac{1}{24}\right)}\,,
\end{equation}
where  $f_n{(k,k^\prime)}$ are polynomials in $k$ and $k'$
\begin{align}
    f_1{(k,k^\prime)} =& \frac{ \pi ^2 u^2}{72} (24 k-1)  \left(24 k'-1\right)+\pi  u \left(k'+k-\frac{1}{12}\right)+\frac{3}{8}\\
    f_2{(k,k^\prime)} =&\frac{\pi ^4 (1-24 k)^2 u^4 \left(1-24 k'\right)^2}{10368}-\frac{ \pi ^3 u^3}{864} (24 k-1)  \left(12 k'+12 k-1\right) \left(24 k'-1\right)\\
    &+\frac{\pi ^2 u^2 }{4} \left(\left(k-k'\right)^2-3 \left(k'+k-\frac{1}{12}\right)^2\right)-\frac{3\pi  u }{8} \left(k'+k-\frac{1}{12}\right)-\frac{15}{128}.
\end{align}
The  $f_n{(k,k^\prime)}$
can be replaced by  differential operators with respect to $\tau$ or $\bar{\tau}$ acting on the exponential 
\begin{equation}
    Z_\text{U} = e^{-\mathcal{A}} \left(\dfrac{\sigma L_z^2}{2\pi u} \right)^{\frac{1}{2}}
    \left(1+\sum_{n=1}^\infty 
    \dfrac{f_n\left(
        \dfrac{\partial_\tau}{2 \pi i},
        -\dfrac{\partial_{\bar{\tau}}}{2 \pi i}\right)}
    {\mathcal{A}^n} 
    \right)Z_0,\label{eq:partition function}
\end{equation}
where 
\begin{equation}
    Z_0\equiv \sum_{k,k^\prime} p(k) p(k^\prime) e^{2\pi i \tau \left(k-\frac{1}{24}\right)}
    e^{-2\pi i \bar{\tau} \left(k'-\frac{1}{24}\right)}=\eta^{-1}(q)\eta^{-1}(\bar{q}).
\end{equation}
$\eta$ is the Dedekind eta function and $q\equiv e^{2\pi i \tau}$. In order to find a closed expression for the partition function at each order in $1/\mathcal{A}$, we require the following properties of the Dedekind eta functions and holomorphic Eisenstein series \cite{apostolModularFunctionsDirichlet1990a}
\begin{align}
    q \partial_q \eta^{-1}(q)&=-\eta^{-1}(q)\dfrac{E_2(q)}{24}\\
    q \partial_q E_2(q)&=\dfrac{E_2(q)^2-E_4(q)}{12}\\
     q \partial_q E_4(q)&=\dfrac{E_2(q)E_4(q)-E_6(q)}{3}\\
     q \partial_q E_6(q)&=\dfrac{E_2(q)E_6(q)-E_4(q)^2}{2}.
\end{align}
Finally, we can compute the large area expansion of the free energy 
\begin{equation}\label{eq:free energy}
   F_\text{U} = -\log Z_\text{U} = \mathcal{A} -\dfrac{1}{2}\log\left(\frac{\sigma i}{\pi  (\tau-\bar\tau)}L_{z}^{2}\right)+2\log\left|\eta\left(\tau\right)\right|+\sum_{n=1}^{\infty}\dfrac{g_{n}(\tau,\bar{\tau})}{\mathcal{A}^{n}}
\end{equation}
with 
\begin{align*}
g_{1}= & -\frac{\pi^{2}\left(\tau-\bar{\tau}\right)^{2}}{288}\left|E_{2}\right|^{2}+\frac{i\pi\left(\tau-\bar{\tau}\right)}{48}\text{Re}E_{2}+\frac{3}{8}
\end{align*}
\begin{align*}
g_{2}= & -\frac{\pi^{4}\left(\tau-\bar{\tau}\right)^{4}}{41472}\left(\text{Re}\left[E_{4}(\tau)E_{2}\left(-\bar{\tau}\right)^{2}\right]-\left|E_{4}\right|^{2}\right)+\frac{i\pi^{3}\left(\tau-\bar{\tau}\right)^{3}}{3456}\text{Re}\left[E_{4}(\tau)E_{2}\left(-\bar{\tau}\right)\right]\\
 & +\frac{\pi^{2}\left(\tau-\bar{\tau}\right)^{2}}{1152}\left(3\left|E_{2}\right|^{2}+\text{Re}E_{4}\right)-\frac{i\pi\left(\tau-\bar{\tau}\right)}{32}\text{Re}E_{2}-\frac{3}{16}
 \end{align*}
 \begin{align*}
g_{3}= & -\frac{\pi^{6}\left(\tau-\bar{\tau}\right)^{6}}{17915904}\left\{ \left|E_{2}\right|^{6}-12\left|E_{2}\right|^{2}\text{Re}\left[E_{2}(\tau)^{2}E_{4}\left(-\bar{\tau}\right)\right]+12\text{Re}\left[E_{2}(\tau)^{3}E_{6}\left(-\bar{\tau}\right)\right]+39\left|E_{4}E_{2}\right|^{2}\right.\\
 & \hfill\qquad\qquad\qquad\quad\left.-72\text{Re}\left[E_{4}(\tau)E_{2}(\tau)E_{6}\left(-\bar{\tau}\right)\right]+32\left|E_{6}\right|^{2}\right\} \\
 & -\frac{i\pi^{5}\left(\tau-\bar{\tau}\right)^{5}}{995328}\left\{ 2\left|E_{2}\right|^{4}\text{Re}E_{2}-12\left|E_{2}\right|^{2}\text{Re}\left[E_{2}(\tau)E_{4}\left(-\bar{\tau}\right)\right]+26\left|E_{4}\right|^{2}\text{Re}E_{2}\right.\\
 & \hfill\qquad\qquad\qquad\quad\left.-4\text{Re}\left[E_{4}(\tau)E_{2}\left(-\bar{\tau}\right)^{3}\right]+12\text{Re}\left[E_{6}(\tau)E_{2}\left(-\bar{\tau}\right)^{2}\right]-24\text{Re}\left[E_{6}(\tau)E_{4}\left(-\bar{\tau}\right)\right]\right\} \\
 & +\frac{\pi^{4}\left(\tau-\bar{\tau}\right)^{4}}{165888}\left\{ 2\left|E_{2}\right|^{2}\text{Re}E_{2}^2(\tau)+2\left|E_{2}\right|^{4}-12\text{Re}E_{4}\left|E_{2}\right|^{2}+12\text{Re}\left[E_{2}(\tau)E_{6}\left(-\bar{\tau}\right)\right]\right.\\
 & \hfill\qquad\qquad\qquad\quad\left.-2\text{Re}\left[E_{4}(\tau)E_{2}\left(-\bar{\tau}\right)^{2}\right]+3\left|E_{4}\right|^{2}\right\} \\
 & -\frac{i\pi^{3}\left(\tau-\bar{\tau}\right)^{3}}{41472}\left(3\left|E_{2}\right|^{2}\text{Re}E_{2}(\tau)+12\text{Re}\left[E_{2}(\tau)E_{4}\left(-\bar{\tau}\right)\right]+\text{Re}E_{2}^{3}-6\text{Re}\left[E_{4}E_{2}\right]+6\text{Re}E_{6}\right)\\
 & +\frac{1}{384}\pi^{2}u^{2}\left(4\left|E_{2}\right|^{2}+2\text{Re}E_{4}\right)-\frac{1}{32}\pi u\text{Re}E_{2}+\frac{21}{128},
\end{align*}
where all the holomorphic Eisenstein series are evaluated at $\tau \equiv \alpha + iu$, unless explicitly stated otherwise. Recalling that $E_{2n}\left(-\frac{1}{\tau}\right)= \tau^{2n} E_{2n}(\tau)$, except for $n=1$ for which $E_2\left(-\frac{1}{\tau}\right)=\tau^{2} E_{2}(\tau)-\frac{6\tau}{\pi i}$, the modular invariance of the above coefficients can be explicitly checked. In app.\;\ref{app:finite Lz} (finite-transverse-volume), we recast the large-area expansion as modular derivatives with respect to $\tau$ and $\bar\tau$ acting on the Gaussian partition function $Z_0(\tau,\bar\tau)$.

This universal partition function  has a Hagedorn transition because the degeneracies grow asymptotically as $\log p(k) \sim \sqrt{k}$, dominating the Boltzmann weight of string modes for sufficiently small areas.
The critical area is:
\begin{equation}
     \mathcal{A}_\text{H} \equiv 
     \begin{cases}
        \dfrac{\pi u}{3} & \text{if } u \geq 1  \vspace{0.1cm} \\
        \dfrac{\pi }{3u} & \text{if } u \leq 1,\\
    \end{cases}
\end{equation}
which matches the no-tachyon condition for the ground state, $\mathcal{E}_{0,0} = \sqrt{1 - \frac{\pi u}{3\mathcal{A}}} \in \mathbb{R}$.
The sum in eq.\;\eqref{eq:partition function} converges quickly, allowing for a precise numerical evaluation even for areas close to this phase transition.

For square domain walls with $\tau=i$, we have the following free energy expansion,
\begin{equation}\label{eq:inverse area expansion of the free energy}
    F_\text{U}=\mathcal{A} + 0.391594...-\dfrac{1}{2}\log\left(\sigma L_{z}^{2}\right)-\frac{0.250000...}{\mathcal{A} }+\frac{0.014107...}{\mathcal{A} ^{2}}+\frac{0.131398...}{\mathcal{A} ^{3}} + \mathcal{O}(\mathcal{A}^{-4}),
\end{equation}
where the numerical coefficients can be computed with arbitrary precision, and grow with the order.

\begin{figure}
    \centering
    \includegraphics[width=1\linewidth]{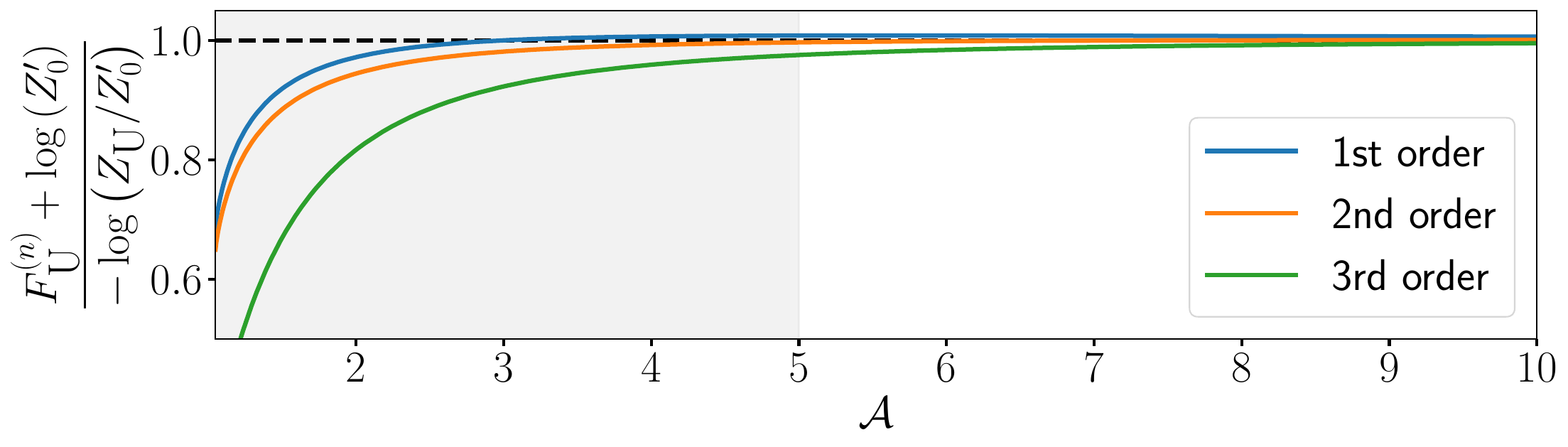}
    \caption{Comparison between the numerical evaluation of eq.\;\eqref{eq:partition function rectangular}, dashed line, and the  expansion  in eq.\;\eqref{eq:free energy}, truncated at order $1/\mathcal{A}^{n}$, full lines. $Z_0^\prime \equiv e^{-\mathcal{A}} \left(\frac{\sigma L_z^2}{2\pi u} \right)^{\frac{1}{2}}Z_0$. The shaded region denotes the parameter range used to extract the value of $\gamma_3$. }
    \label{fig:comparision numerical evaluation of series with the perturbative expansion}
\end{figure}

In figure \ref{fig:comparision numerical evaluation of series with the perturbative expansion}, we compare the universal part of the free energy $F_{\rm{U}}$ with its expansion in $1/\mathcal{A}$. This is an asymptotic expansion. As the plot shows, in the region where we can obtain sufficiently precise Monte-Carlo measurements (see next section), the asymptotic expansion in $1/\mathcal{A}$ is not a good approximation to $F_{\rm{U}}$.
Therefore, in the numerical analysis of the next section, we shall use the full expression \eqref{eq:partition function rectangular} instead of its expansion in $1/\mathcal{A}$.

\subsection{Leading non-universal contribution
}
In this subsection we review the path integral description of the worldsheet, derive the effective field theory in the static gauge \cite{dubovskyEffectiveStringTheory2012, aharonyEffectiveStringTheory2012b}, and compute the leading non-universal contribution for finite worldsheets. The result for long and thin worldsheets is known (see, for example, \cite{aharonyEffectiveActionConfining2009,aharonyEffectiveTheoryLong2011}).

\paragraph{Review of the path integral approach}\label{sec:review path integral}
\hfill %

The leading term in the EST action is 
proportional to the domain wall area, 
known as the Nambu-Goto action:
\begin{equation}
    S_{\rm NG}\left[X\right]=\sigma\int d^{2}\xi\sqrt{\det h}.\label{eq:NG}
\end{equation}
Here, $h_{ab}$ represents the induced metric on a 2-dimensional manifold parametrized  by $\xi = \left(\xi_1,\xi_2\right)\in[0,L_1]\times[0,L_2]$
\begin{equation}
    h_{ab}(\xi) = \partial_a X^\mu(\xi) \partial_b X^\nu(\xi) G_{\mu\nu}(X(\xi)),
\end{equation}
where $X^\mu$ denotes the target space coordinates of the worldsheet, and $G_{\mu\nu}(\xi)$ is the target space metric, which we take to be the Euclidean metric.
We focus on a non-contractible worldsheet configuration arising from the periodic lattice system. By going to the static gauge $X^\mu=\left(\xi^{1},\xi^{2}, X\left(\xi^{1},\xi^{2}\right)\right)$, the Nambu-Goto action becomes
\begin{equation}
    S=\sigma\int_{[0,L_1]\times[0,L_2]}d^{2}\xi\left(1+\left(\partial_{1}X\right)^{2}+ \left(\partial_{2}X\right)^{2}\right)^{1/2}.
\end{equation}

We introduce a counting parameter for the effective field string theory by going to dimensionless coordinates and fields.
The coordinate rescaling, $\xi\to\xi^\prime$, must preserve the aspect ratio and modify the volume form as $d^{2}\xi\to A d^{2}\xi^{\prime}$, where $A=L_1 L_2$.
The field $X$ has dimensions of length, therefore $\pi\equiv\sqrt{\sigma}X$ is dimensionless. Then, we can expand the square root to derive the effective action, composed of: a classical term, the minimal area; the action of a free massless boson canonically normalized; and an infinite set of interactions organized in inverse powers of $\mathcal{A}\equiv A \sigma$
\begin{equation}
    S_\text{NG} = \mathcal{A} +\int_{[0,\sqrt{u}]\times[0,\nicefrac{1}{\sqrt{u}}]}d^{2}\xi \left( \dfrac{1}{2}\partial_i\pi \partial^i \pi +  V\left[\pi\right]\right),
\end{equation}
where $ V\left[\pi\right] = \sum_{n>1} (\mathcal{A})^{-n}  \left(\frac{1}{2}\right)^{(n)} (\partial_i \pi \partial_i \pi)^n$, and $x^{(n)}$ is the factorial power.

We consider the Gaussian theory within a finite worldsheet volume, but in an infinite target space (i.e., $L_1$ and $L_2$ are finite while $L_z$ is infinite). 
Its propagator is
\begin{equation}
G\left(\xi^{1},\xi^{2};\xi^{1\prime},\xi^{2\prime}\right)=\frac{1}{4\pi}\sum_{ \begin{subarray}{c}m,n=-\infty\\ (m,n)\neq(0,0)\end{subarray}}^{+\infty}\frac{e^{\left(2\pi im/L_{1}\right) \left(\xi^{1}-\xi^{1\prime}\right)}e^{\left(2\pi in/L_{2}\right)\left(\xi^{2} -\xi^{2\prime}\right)}}{\pi\left|m+iun\right|^{2}}u.
\label{eq:propagator}
\end{equation}
Notice that the zero mode, \((m,n) = (0,0)\), is absent. 
This mode %
corresponds to translations in the \(z\) direction and   it has already been integrated out in eq.\;\eqref{eq:partition function}, resulting in the entropic factor of $L_z$. 
As all fields are paired with a derivative, we do not expect any perturbative contributions from the zero-mode.

We review the computation of the Gaussian partition function using zeta function regularization in app.\;\ref{sec:Gaussian partition function} and confirm the normalization in eq.\;\eqref{eq:partition function}.

\paragraph{Leading non-universal contribution}\label{subsec:non-universal contribution}\hfill %

We now consider the first non-universal contribution to effective string theory, arising from the extrinsic curvature, $K^\mu_{ij} \equiv \nabla_i \partial_j X^\mu$, to the fourth power. 
In the static gauge, this contribution leads to multiple operators with varying powers of $\mathcal{A}$. The lowest-order term, at order 3, is
\begin{equation}
    V_{\text{NU}}\left[\pi\right]= %
    -2\dfrac{\gamma_{3}}{\mathcal{A}^{3}}\int d^{2}\xi\left(\partial_{i}\partial_{j}\pi\partial^{i}\partial^{j}\pi\right)^{2},
\end{equation}

The leading non-universal contribution to the free energy of finite worldsheets, computed to first order in $\gamma_3$ and order 3 in $1/\mathcal{A}$, computed in app.\;\ref{subsec:green functions} following \cite{dietzRenormalizationStringFunctionals1983}, is
\begin{align}
    F_{\text{NU}}(iu)&=-\dfrac{32\gamma_{3} \pi^{6}}{225 \mathcal{A}^{3}}u^{4}E_{4}\left(iu\right)^2
+O(\mathcal{A}^{-4})\label{eq:non universal contribution for no twist}\\
    F_{\text{NU}}(i)&=-\dfrac{\gamma_{3}}{\mathcal{A}^{3}}\frac{\Gamma \left(\frac{1}{4}\right)^{16}}{3200 \pi ^6}+O(\mathcal{A}^{-4})\,,\label{eq:non universal contribution for square domain walls}
\end{align}
where the last equation shows its evaluation at $u=1$, using the results for the Eisenstein series in \cite{10.1112/blms/bdn014}, which corresponds to a square worldsheet. 

Eq.\;\eqref{eq:non universal contribution for no twist} can be checked by matching the ground state energy in the long string limit, $L_1\to\infty$ at fixed $L_2$ or equivalently $u\to\infty$ and $\mathcal{A}\to\infty$ at fixed ratio. It reproduces the known correction to the ground state energy \cite{miroFluxTubeSmatrix2019,Aharony:2010db}
\begin{equation}
    \Delta E_0\equiv    \lim_{L_{1}\to\infty}\dfrac{\Delta F_{\text{NU}}(L_1,L_2)}{L_{1}}=-\dfrac{32\pi^{6}}{225}\dfrac{\sqrt{\sigma} \gamma_{3}}{(\sqrt{\sigma} L_{2})^{7}}+\dots
\end{equation}

It would be interesting to compute the non-universal contribution to the free energy at the next order: $O(1/\mathcal{A}^4)$. In fact, this is fully determined by $\gamma_3$ because the next Wilson coefficient starts to contribute at order $1/\mathcal{A}^5$. 
Unfortunately, the zeta-function regularization method we used becomes more difficult at the next order. Firstly, the sums that require regularization, coming from loops, are more complex, like for example
\begin{equation}
\prod_{i=1}^{4}\sum_{\begin{subarray}{c}m_i,n_i=-\infty\\ (m_i,n_i)\neq(0,0)\end{subarray}}^{+\infty}\frac{m_i^{a_i}n_i^{b_i}}{|m_i+iun_i|^{2}}\delta_{\sum_i m_i,0}\delta_{\sum_i n_i,0}\,.
\end{equation}
Secondly, there are counterterms that are forbidden by the non-linearly realized bulk Lorentz symmetry but that are generated in the zeta-function regularization scheme \cite{dubovskyEffectiveStringTheory2012}.

\subsection{subleading non-universal contribution}\label{subsec:subleading TBA}
Based on the discussion above, a different approach is required. The argument in subsec.~\ref{subsec: universal contribution} for computing the partition function from the GGRT spectrum generalizes to an arbitrary worldsheet spectrum as follows. If the string rest mass is
\begin{equation}
m \equiv \sigma L_2\,\mathcal{E}_I,
\end{equation}
with \(\mathcal{E}_I\) the energy density determined by worldsheet excitations labeled by \(I\), then eq.~\eqref{eq:qm ggrt spectrum} becomes
\begin{equation}
E_I=\sigma L_2\,\mathcal{E}_{I,p_z}
=\sigma L_2\sqrt{\mathcal{E}_I^{\,2}+\left(\frac{p_z}{\sigma L_2}\right)^2}\, .
\end{equation}
Integrating over the transverse volume as before yields
\begin{equation}\label{eq:generic partition function}
Z=\frac{\sqrt{\mathcal{A}\,\sigma L_z^{2}}}{\pi\sqrt{u}}
\sum_I p(I)\,\mathcal{E}_I\,K_1\!\left(\mathcal{A}\,\mathcal{E}_I\right),
\end{equation}
where \(p(I)\) is the degeneracy of the quantum number \(I\). In \(d\) target-space dimensions this generalizes to
\begin{equation}
    Z=\frac{V_T}{(2\pi)^{\,d-2}}\sum_I p(I)\!\int_{-\infty}^{+\infty}\! d^{\,d-2}p\;
e^{-\sigma L_1L_2\,\mathcal{E}_{I,p}}
=\sqrt{\frac{2\mathcal{A}}{\pi}}
\left(\frac{\sigma}{2\pi u}\right)^{\!\frac{d-2}{2}}V_T
\sum_I p(I)\,K_{\frac{d-1}{2}}\!\left(\mathcal{A}\mathcal{E}_I\right)\mathcal{E}_I^{\frac{d-1}{2}},
\end{equation}
in agreement with \cite{billoPartitionFunctionInterfaces2006,berattoDeformationCompactifiedBoson2020a}, up to normalization.

Next, we compute the spectrum at leading order in \(\gamma_3\). Using the low-energy integrability of the \(d=3\) worldsheet theory \cite{chenUndressingConfiningFlux2018,Cooper:2014noa,dubovskySolvingSimplestTheory2012}, we employ the Thermodynamic Bethe Ansatz (TBA) to obtain  the excited-state energy densities  (see app.~\ref{app:TBA})
\begin{equation}
\mathcal{E}_{k,k',s,s'}=\mathcal{E}_{k,k'}
-\gamma_3\underbrace{\frac{2048\pi^6 u^4}{225\mathcal{A}^3}
\frac{(240s+1)(240s'+1)}{\mathcal{E}_{k,k'}
\left[\left(\mathcal{E}_{k,k'}+1\right)^2-\frac{\pi^2u^2}{\mathcal{A}^2}\left(k-k'\right)^2\right]^{3}}}_{-\Delta\mathcal{E}_{k,k',s,s'}},
\end{equation}
where \(\mathcal{E}_{k,k'}\) is the GGRT spectrum (eq.~\eqref{eq:GGRT spectrum}), \(k\in\mathbb{N}_0\), and \(s\) is the sum of cubes over a partition of \(k\) (and likewise for \(s'\), \(k'\)). The TBA result is valid up to order $\mathcal{O}(1/\mathcal{A}^5)$, where $\gamma_3^2$-induced inelasticity arises.

Expanding eq.~\eqref{eq:generic partition function} to first order in \(\gamma_3\) gives
\begin{equation}\label{eq:full partition function}
Z=\frac{\sqrt{\mathcal{A}\sigma L_z^{2}}}{\pi\sqrt{u}}
\sum_{k,k',s,s'} p(k,s)p(k',s')\Bigg(\mathcal{E}_{k,k'}K_1\left(\mathcal{A}\mathcal{E}_{k,k'}\right)
-\mathcal{A}\gamma_3 \Delta\mathcal{E}_{k,k',s,s'}\mathcal{E}_{k,k'}K_0\left(\mathcal{A}\mathcal{E}_{k,k'}\right)\Bigg)
+\mathcal{O}(\gamma_3^2),
\end{equation}
where the first term reduces to the universal partition function since \(\sum_s p(k,s)=p(k)\).

The second term, which we denote as $\gamma_3 Z_{\gamma_3}$, 
\begin{equation}\label{eq:definition Zg3}
    Z_{\gamma_3} \equiv-\frac{\sqrt{\sigma L_z^{2}}}{\pi\sqrt{\mathcal{A}u}} \sum_{k,k',s,s'} p(k,s)p(k',s') \Delta\mathcal{E}_{k,k',s,s'}\mathcal{E}_{k,k'}K_0\left(\mathcal{A}\mathcal{E}_{k,k'}\right)
\end{equation}
can be computed as in subsec.\;\ref{subsec: universal contribution}, by expanding at large area and replacing the powers of $k$ and $k^\prime$ with derivatives with respect to $\tau$ and $\bar\tau$ acting on $e^{2\pi i\tau\left(k-1/12\right)}e^{-2\pi i\bar{\tau}\left(k^{\prime}-1/12\right)}$. This gives
\begin{align*}
Z_{\gamma_3}\left(\tau=\alpha+iu\right) & =\dfrac{32\pi^{6}u^{4}}{225\mathcal{A}^{3}}\dfrac{\sqrt{\sigma L_{z}^{2}}}{\sqrt{2\pi u}}e^{-\mathcal{A}}\sum_{n=0}\dfrac{t_{n}\left(\dfrac{\partial_{\tau}}{2\pi i}+\dfrac{1}{24},-\dfrac{\partial_{\bar{\tau}}}{2\pi i}+\dfrac{1}{24}\right)}{\mathcal{A}^{n}}\\
 & \times\underbrace{\sum_{k,k^{\prime},s,s^{\prime}}(240s+1)p\left(k,s\right)(240s^{\prime}+1)p\left(k^{\prime},s^{\prime}\right)e^{2\pi i\tau\left(k-1/12\right)}e^{-2\pi i\bar{\tau}\left(k^{\prime}-1/12\right)}}_{Z_{\gamma_{3}}^{0}\left(\tau\right)},
\end{align*}
with
\begin{align*}
t_{0} & =1\\
t_{1} & =\frac{1}{72}\pi^{2}\left(24k-1\right)u^{2}\left(24k'-1\right)-7\pi u\left(k'+k-\frac{1}{12}\right)-\frac{1}{8}\\
t_{2} & =\cdots . 
\end{align*}

 The generating function for the leading \(\gamma_3\) piece is computed in app.\;\ref{app:z0gamma3}, and it is 
\begin{equation}
    Z^0_{\gamma_3}(\tau)=\left|\frac{E_4(\tau)}{\eta(\tau)}\right|^2,
\end{equation}
thus the leading correction reads
\begin{equation}
Z_{\gamma_3}
=\frac{32\pi^6u^4}{225\,\mathcal{A}^3}\,
\frac{\sqrt{\sigma L_z^{2}}}{\sqrt{2\pi u}}\,e^{-\mathcal{A}}\left(
\left|\frac{E_4(\tau)}{\eta(\tau)}\right|^{2} + \mathcal{O}(\mathcal{A}^{-4})\right)
\end{equation}
It agrees with the path-integral computation for zero twist (\(\tau=iu\)), eq.~\eqref{eq:non universal contribution for no twist}, and extends it to generic twist \(\tau=\alpha+iu\).

The extension to the odd-$n$ Wilson coefficients $\gamma_n$ (the coefficient of $s^n$  in the $2\!\to\!2$ phase shift) is straightforward, but meaningful only in the regime where the phase shift remains real (i.e. before integrability is lost, effectively only $n=3,5,7$). The general result is:

\begin{equation}
    \Delta \mathcal{E}_{k,k^\prime,s,s^\prime} = -2^{6n-1}\pi^{2n}\zeta(-n)^{2}\frac{\left(1+\dfrac{2}{\zeta(-n)}s\right)\left(1+\dfrac{2}{\zeta(-n)}s^{\prime}\right)}{R^{2+2n}\mathcal{E}_{k,k^{\prime}}\left(\left(\mathcal{E}_{k,k^{\prime}}+1\right)^{2}-\dfrac{4\pi^{2}\left(k-k^{\prime}\right)}{R^{4}}\right)^{n}},
\end{equation}
where $s$ is now given by $\sum_l n_l^n$ (with $\sum_l n_l=k$), and the correction to the partition function is
\begin{equation}
    \gamma_{n}Z_{\gamma_n}=-\dfrac{1}{2}\left(4^{n}\pi^{n}\zeta(-n)\right)^2\dfrac{\sqrt{\sigma L_{z}^{2}}}{\sqrt{2\pi u}}e^{-\mathcal{A}}\dfrac{u^{n+1}}{\mathcal{A}^{n}}\gamma_{n}\left(\left|\dfrac{E_{n+1}\left(\tau\right)}{\eta\left(\tau\right)}\right|^{2} + {\cal O} ({\cal A}^{-n-1})\right).\label{eq:higher order corrections.}
\end{equation}

The subleading \(\gamma_3\) contribution is obtained by acting with the derivatives on \(Z^0_{\gamma_3}\), yielding
\begin{align}\label{eq:Zg3 expansion}
Z_{\gamma_3}=\dfrac{32\pi^{6}}{225\mathcal{A}^{4}}\dfrac{\sqrt{\sigma L_{z}^{2}}}{\sqrt{2\pi u}}\dfrac{e^{-\mathcal{A}}}{|\eta(\tau)|^2}\Bigg[& u^4|E_4|^2 \mathcal{A} +\frac{1}{72}\pi^{2}u^{6}\left|7E_{2}E_{4}-8E_{6}\right|^{2}\\
 & +\frac{7}{12}\pi u^{5}\left(8\text{Re}\left(E_{4}(\tau)E_{6}\left(-\bar{\tau}\right)\right)-7\left|E_{4}\right|^{2}\text{Re}\left(E_{2}\right)\right)\notag\\
 & -\frac{1}{8}u^{4}\left|E_{4}\right|^{2} + \mathcal{O}(1/\mathcal{A})\Bigg],\notag
\end{align}
where all the holomorphic Eisenstein series are evaluated at $\tau \equiv \alpha + iu$, unless explicitly stated otherwise.

Writing the free energy $F=-\log Z = -\log (Z_\text{U}+\gamma_3 Z_{\gamma_3}+\mathcal{O}(\gamma_3^2))=F_\text{U} +\gamma_3 F_{\gamma_3}+\mathcal{O}(\gamma_3^2)$, we obtain 
\begin{equation}
    F_{\gamma_3} \equiv -\dfrac{Z_{\gamma_3}}{Z_\text{U}} = \mathcal{A}\dfrac{\sum_{k,k',s,s'}p(k,s)p(k',s')\Delta\mathcal{E}_{k,k',s,s'}\mathcal{E}_{k,k'}K_{0}\left(\mathcal{A}\mathcal{E}_{k,k'}\right)}{\sum_{k,k'}p(k)p(k')\mathcal{E}_{k,k'}K_{1}\left(\mathcal{A}\mathcal{E}_{k,k'}\right)},
    \label{def:Fg3}
\end{equation}
which can be efficiently evaluated numerically. For the simulated worldsheets, with $\tau=i$, we obtain 
\begin{equation}\label{eq:free energy perturbatively in gamma3}
    F_{\gamma_3} = \frac{32\pi^6}{225\,\mathcal{A}^3}E_4(i)^2 \left(1- \dfrac{13}{2 \mathcal{A}} + \mathcal{O}(\mathcal{A}^{-2})\right).
\end{equation}

\begin{figure}[t]
    \centering
    \includegraphics[width=1\linewidth]{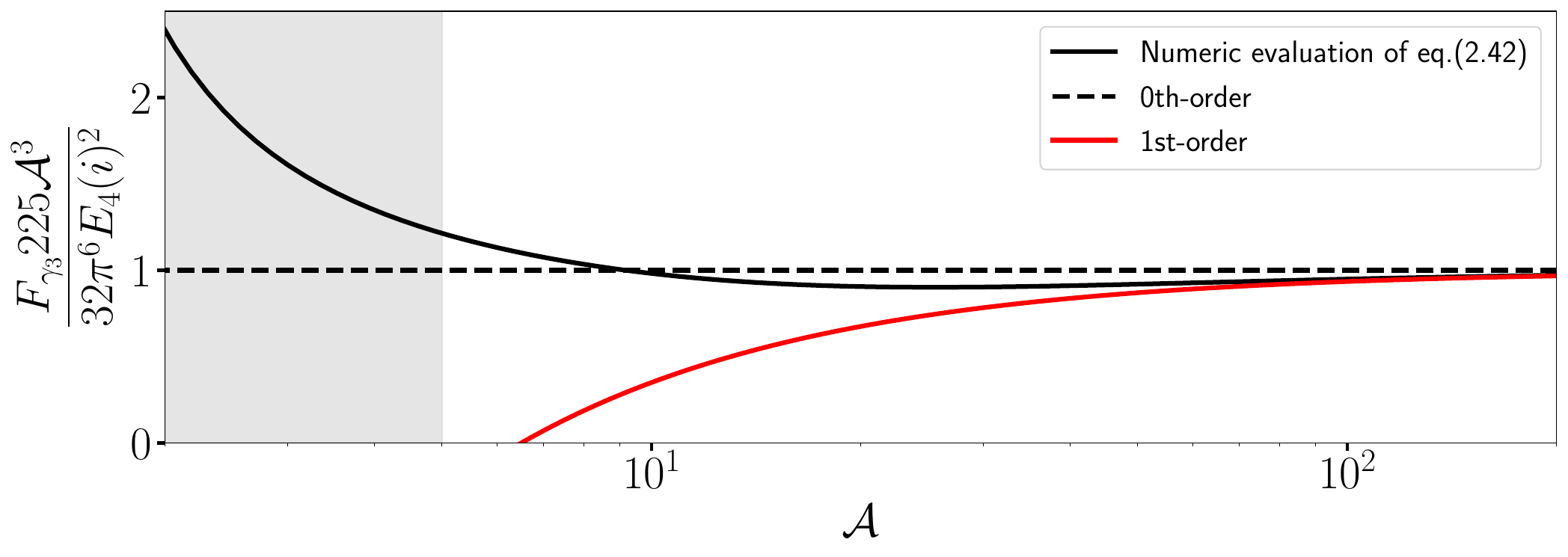}
    \caption{Comparison of the exact $F_{\gamma_3}$, eq.\;\eqref{def:Fg3}, with its asymptotic expansion, eq.\;\eqref{eq:free energy perturbatively in gamma3}. 
    The gray region marks the small $\mathcal{A}$ range relevant to our numerics, where the two differ by as much as a factor of two. Left of the zero, the 1st-order prediction has the wrong sign.
    }
    \label{fig:comparison series to exact Zgamma_3}
\end{figure}

For moderate or small $\mathcal{A}$ the optimal truncation is the leading term. In fig.\;\ref{fig:comparison series to exact Zgamma_3} we compare the large area expansion with the exact $F_{\gamma_3}$, obtained from evaluating eq.\;\eqref{def:Fg3} numerically; the shaded region indicates the range of $\mathcal{A}$ where $\gamma_3$ effects are detectable within the numeric precision of the Monte-Carlo simulations described in the next section. The zeroth-order term alone does not reproduce the exact result, and adding the first correction worsens the agreement in the relevant region. Because our simulations probe the small-$\mathcal{A}$ regime, we use the exact expression \eqref{def:Fg3}  rather than its large-area expansion. %

\begin{figure}
    \centering
    \includegraphics[width=0.32\linewidth]{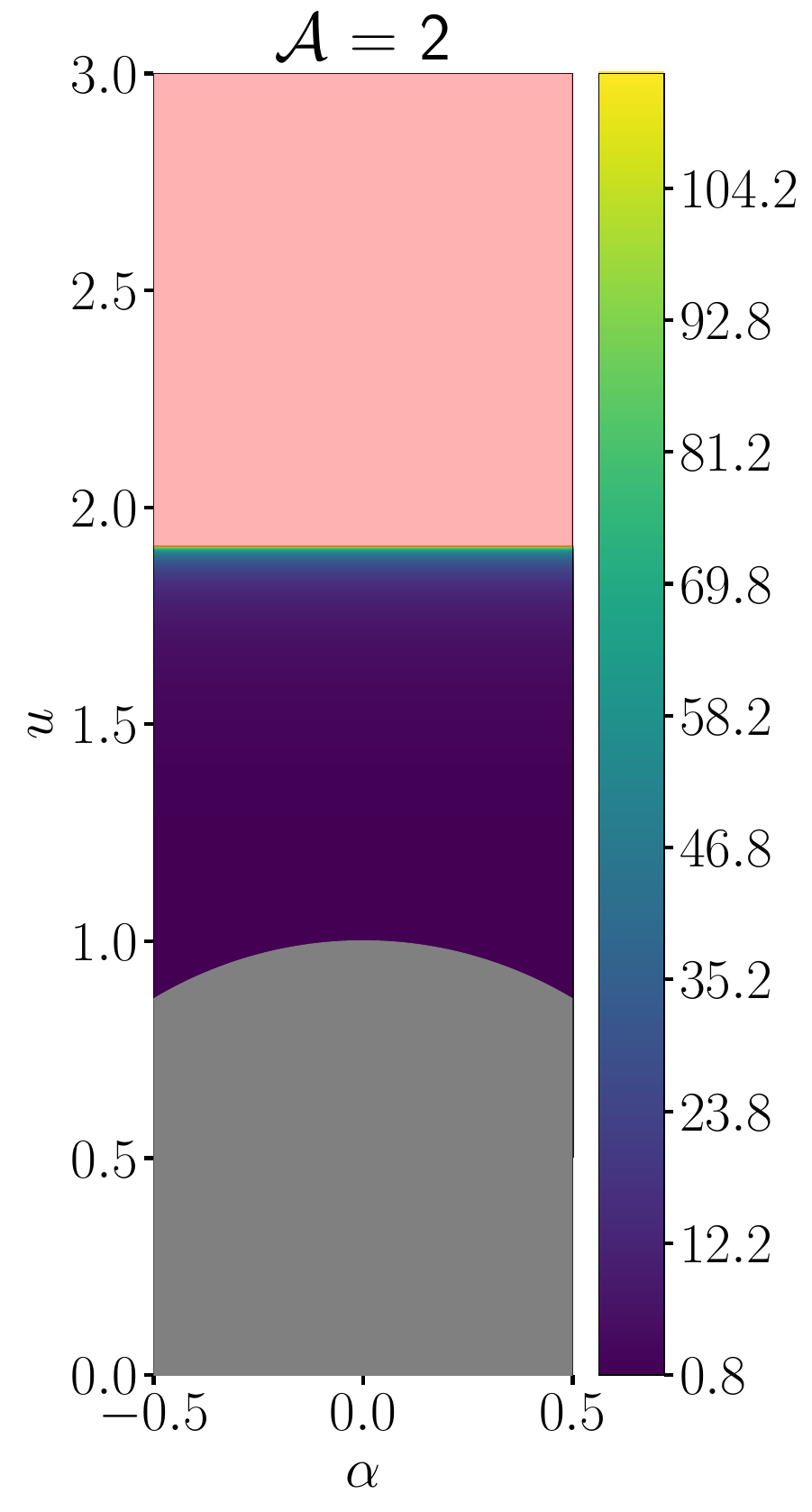}\includegraphics[width=0.32\linewidth]{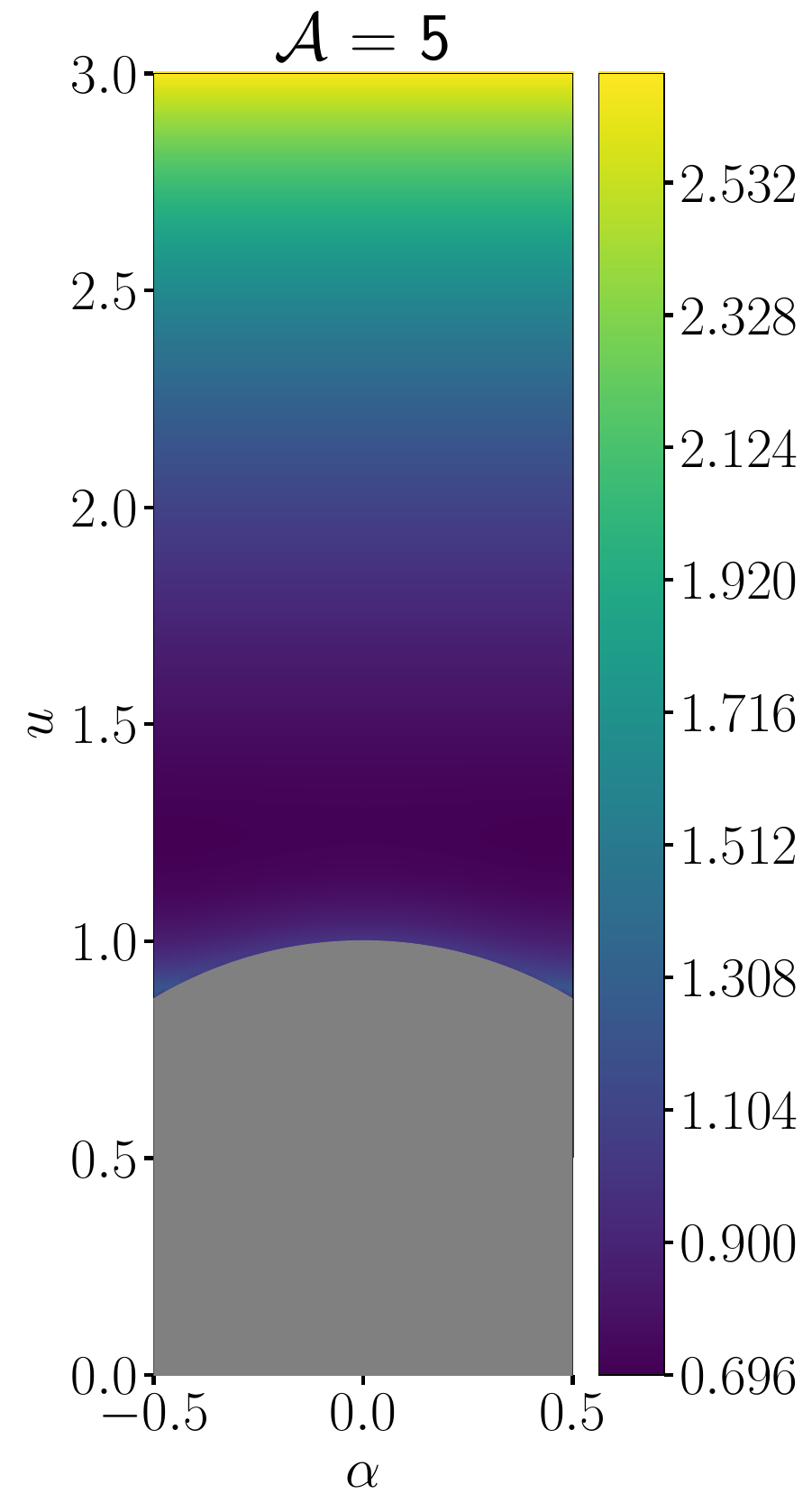}\includegraphics[width=0.32\linewidth]{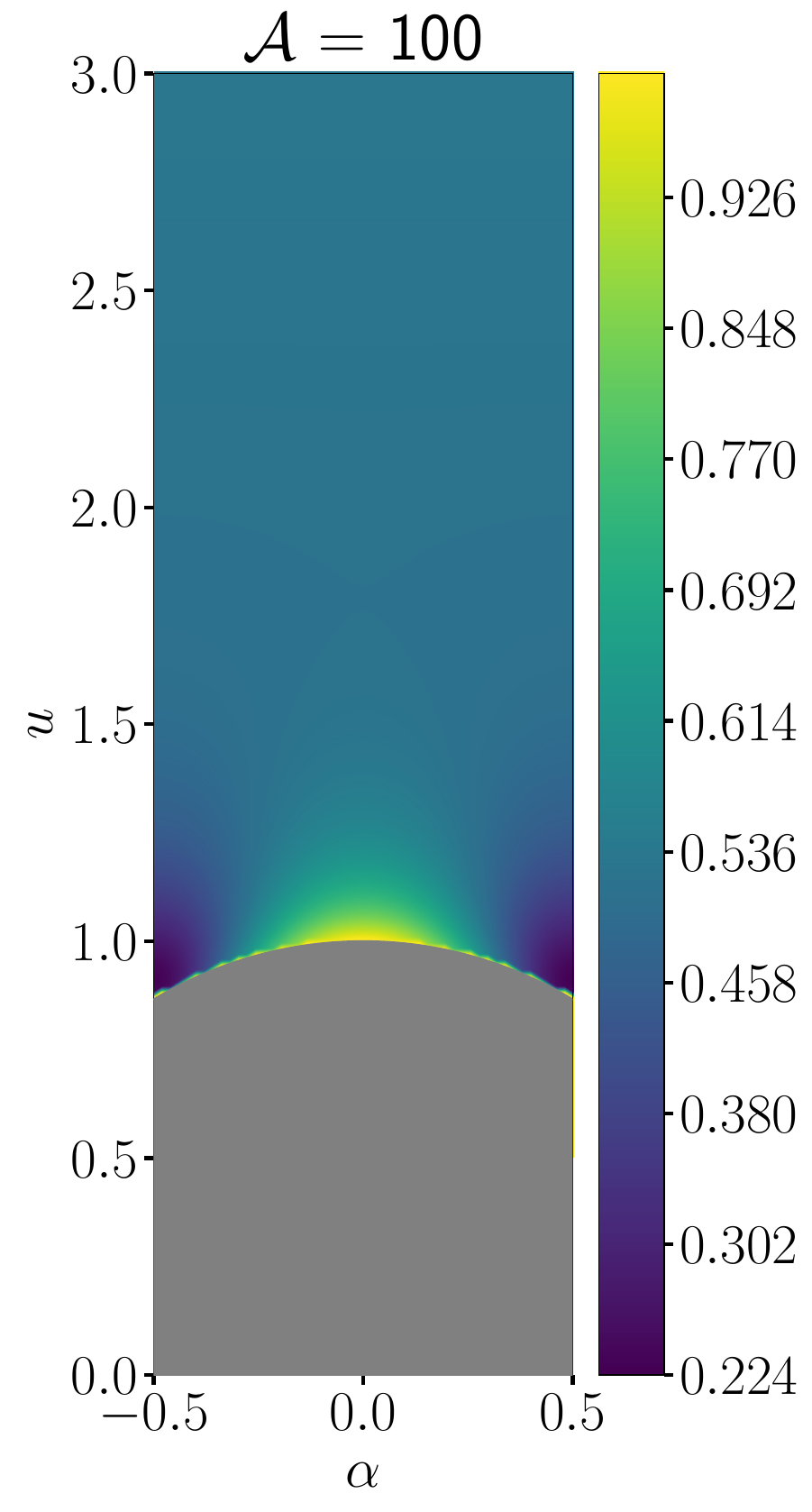}
    \caption{Dependence of the ratio $\nicefrac{F_{\gamma_3}}{F_\text{U}}$ %
    between the leading non-universal correction proportional to $\gamma_3$ and the universal free-energy, as a function of the modular parameter of the torus for 3 representative areas. The red region for ${\cal A}=2$ is beyond the Hagedorn phase transition due to the smallest size when we change the aspect ratio at fixed area. The ratio is normalized by the value at $\tau=i$.}
    \label{fig:dependence in the modular parameter}
\end{figure}
\paragraph{Exploring the fundamental domain} \hfill \break
Having computed both the universal and non-universal contributions for generic interfaces, we now investigate whether there exist points in the fundamental domain where the non-universal terms are enhanced relative to the universal ones. In fig.\;\ref{fig:dependence in the modular parameter}, we present the ratio between the non-universal contributions at first order in $\gamma_3$ and the universal part, $\nicefrac{F_{\gamma_3}}{F_\text{U}}$, within the first fundamental domain up to $u=3$. The analysis is performed for three representative areas: (i) the smallest area at which a determination of $\gamma_3$ is still feasible despite contaminations from other sources; (ii) the largest area for which we have sufficient statistics to measure $\gamma_3$; and (iii) the largest area typically employed to extract the string tension. The overall conclusion is that the dependence on the twist parameter $\alpha$ is weak for the areas where $\gamma_3$ can be reliably measured. The dependence on $u$, however, is more pronounced and suggests that squeezing the interface could improve the extraction of $\gamma_3$. This advantage comes at a cost: simulations become more challenging, as one must ensure that both lattice sizes remain much larger than the correlation length, thus narrowing the range of areas where the extraction of $\gamma_3$ remains reliable.
As an alternative, $\gamma_3$ may be determined through a joint fit in the area and the ratio.

\section{Simulation setup for 3d Ising}\label{sec:simulation set up}

The effective string theory worldsheet is realized as a domain wall in the 3d Ising model, arising from the spontaneous symmetry breaking of $\mathbb{Z}_2$.
In the broken phase, but near the bulk critical point, where continuous symmetries are restored, a domain wall is generated by imposing anti-periodic boundary conditions along a plane.
Since bulk physics away from the domain wall is unaffected by the anti-periodic boundary conditions, the ratio of the partition functions with anti-periodic and periodic boundary conditions is solely related with the free energy of the domain wall.

Within the Monte Carlo literature there are many algorithms, each with tradeoffs. The standard single–spin–flip Metropolis sampler of the canonical ensemble \cite{Metropolis:1953am,Hastings:1970aa}, although broadly applicable, is too inefficient for our system. We therefore turn to extended or modified dynamics and ensembles. Before delving into that, let us briefly mention that this system has a rich history starting long before the recent revival of effective string–inspired studies of domain walls. Thus, many numerical methods have been considered (e.g.\ \cite{hasenbuschDirectMonteCarlo1993} standard Monte Carlo, \cite{hasenbuschinterfaceTension3dimensional1997} integration method, \cite{caselleStringEffects3d2003} reweighting, \cite{caselleStaticQuarkPotential2004} multi-level, \cite{caselleEffectiveStringSpectrum2006} multispin and microcanonical demon-update, \cite{caselleJarzynskisTheoremLattice2016} Jarzynski). 

Here we focus on methods that modify or extend the sampled ensemble, which provide greater flexibility and performance. Such extended ensembles underpin extended-canonical/microcanonical and flat-histogram strategies that clarify the order of phase-transition diagnostics and enable extraction of microcanonical entropies in thermodynamically unstable regions \cite{PhysRevE.74.046702,Florio:2021uoz}; tailored update schemes likewise enable precise measurements in extreme sectors, e.g. large charge in the critical O(2) model \cite{Cuomo:2023mxg}. For this model, we developed an improved flat-histogram procedure, inspired by the multicanonical approach \cite{bergMulticanonicalEnsembleNew1992a} and reviewed in \cite{bergMulticanonicalMonteCarlo2002a}, which we use to measure the free energy as a function of the boundary coupling.

Ideally, one would simulate both systems independently and compute the ratio of their partition functions. However, this is not feasible as normalization of partition functions cannot be ensured.
To resolve this, we expand the configuration space so that different boundary conditions become continuously connected in this extended space. Specifically, we allow the coupling in a plane to vary  between $-1$ (antiperiodic) and $1$ (periodic). More precisely, we consider the energy functional
\begin{equation}
H[\{s\};J]=\sum_{\langle i,j \rangle} J_{ij} s_i s_j,
\end{equation}
where $J_{ij} = 1$ if the bond $i,j$ does not cross the $z$-plane at $z = L_z$, and $J_{ij} =J$ if it does. 
The partition function is given by
\begin{equation}
Z\left[\beta\right]	=\int dJe^{-F\left(J\right)} \qquad \qquad  e^{-F\left(J\right)}\equiv\sum_{\left\{ s_{i}\right\} }e^{-\beta H\left[\left\{ s_{i}\right\} ;J\right]}\,.
\end{equation}
Then $e^{-F(1)} = Z_\text{P}$ and $e^{-F(-1)} = Z_\text{AP}$.

To flatten the histogram, we add a weight $\omega(J)$ to the action
\begin{equation}
Z\left[\beta\right]	=\int dJe^{-F\left(J\right)+\omega(J)}, 
\end{equation}
and build $\omega(J)$ to approximate $F(J)$. The weight can be estimated with standard methods such as  Wang-Landau \cite{PhysRevLett.86.2050}. Then, the logarithm of the ratio between the partition functions is given by $\omega(1) - \omega(-1)$.
We used 512 bins by default but dynamically increase the number if the difference between $\omega(J)$ in consecutive bins becomes too large. 
We employ the Wolff algorithm \cite{Wolff:1988uh} for spin updates and a Metropolis update for the boundary coupling.

We sampled multiple inverse temperatures $\beta\in\{0.223102,0.224,0.227,0.2285,0.23\}$. The further away from the critical temperature  $\beta_c = 0.221654626(5)$ \cite{ferrenbergPushingLimitsMonte2018}, the larger the physical volume of the lattice we can sample, or alternatively, the cheaper is to sample the same physical volume (\emph{i.e.} volume in units of the correlation length). Move too far away from the critical temperature and  lattice effects become relevant. $\beta=0.23$ was the maximum inverse temperature we could use before observing significant deviations to the expected continuum behaviour. In the plots we mostly show the data for $\beta=0.23$ as it is the temperature for which we have the largest coverage. Nonetheless, this is representative of the raw data obtained for all other inverse temperatures. We also consider $\beta=0.222$. However, we were unable to go to lattices large enough to observe the linear behavior of the free energy, hence we excluded them from the analysis. 

 We explored whether different aspect ratios and/or twists of the torus could improve the ratio $F_\text{NU}/F_\text{U}$.
While this indeed happens, it comes at the cost of increased simulation time due to the larger physical volumes required to measure $\gamma_3$.
The optimal modular parameter remains undetermined.

The non-universal contributions to the free energy are significantly smaller than the universal ones, necessitating tight control over systematic errors.
Before presenting our results, we address the challenges encountered and the strategies employed to mitigate them.
We identified three primary issues: ignorance of the microscopic details of the lattice configurations (discussed in subsec.\;\ref{subsec:dilute gas}); extremely large diffusion times due to the high autocorrelation of the Monte Carlo dynamics when updating the boundary coupling; and the discretized nature of the $\omega(J)$.
The first issue is well-known and widely discussed, while the latter two are specific to the flat-histogram methods.
We address these issues in the following paragraphs.

\begin{figure}[t]
    \centering
    \includegraphics[width=1\linewidth]{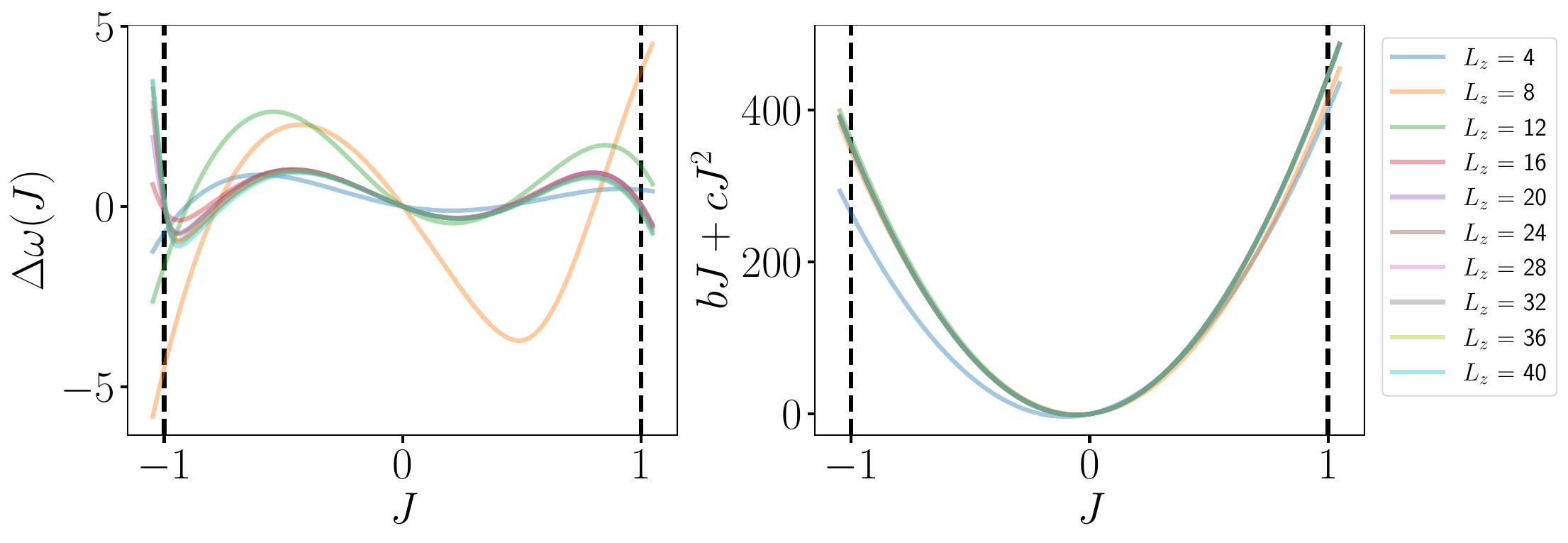}
    \caption{Discrete \textbf{(left)} and continuous \textbf{(right)} contribution to $\omega(J)$ for square domain walls of size $64^2$ and $\beta=0.23$, with 512 bins.}
    \label{fig:multicanonical weight}
 \vspace{-0.5cm}\end{figure}

\paragraph{Range of the boundary coupling}\hfill 

Extending the boundary coupling range beyond $[-1, 1]$ is a natural approach to improving the measurement of $\omega(1)$ and $\omega(-1)$.
However, when the coupling crosses $-1$, there is a "transition": for $J > -1$, the domain wall is delocalized in the bulk but repelled from the plane with variable coupling, while for $J < -1$, the domain wall localizes to the plane with variable coupling.
In the latter regime, the autocorrelation time of the Monte Carlo dynamics increases significantly.
In practice, we used $J \in [-1.05, 1.05]$ for small systems and/or away from the bulk phase transition, but restricted to $J \in [-1.01, 1.01]$ otherwise.
A hint of this behavior can be observed near $J=-1$ in the left panel of fig.~\ref{fig:multicanonical weight}, where the curves with $L_z \geq 20$ exhibit larger variations in that region than in the rest of the domain.

\paragraph{Improved flat-histogram }\hfill 

The main cost is constructing the weight required to flatten the histogram over $J$; because free energy is extensive, the cost grows with system size.
Moreover, the standard flat-histogram method approximates this weight by a piecewise constant function, the logarithm of a histogram. 
Hence, the error is roughly proportional to the weights' variation within each bin.
There are two remedies: (1) increase the number of bins, at the cost of increased simulation time, or (2) approximate part of the weight with a continuous function
\begin{equation}
\omega(J)=\underbrace{bJ+cJ^{2}}_{\text{continuous}}+\underbrace{\Delta\omega(J)}_{\text{discrete}}.
\end{equation}
This provides a middle ground between entropic sampling (as in the Wang--Landau algorithm \cite{PhysRevLett.86.2050}) and the fully continuous piecewise approximations in \cite{PhysRevE.74.046702}.
Then, as long as $\Delta\omega$ varies slowly within each bin, it does not matter how much the weight actually varies.
We ignore the irrelevant constant term, as the weight is defined up to an overall constant.

Although this method may not universally work, for our system, the continuous contribution is often larger then the discrete part by two orders of magnitude, as shown in fig.\;\ref{fig:multicanonical weight}. This drastically reduces the systematic errors in $\omega$, at negligible extra computational cost.

\begin{figure}[t]
    \centering
    \includegraphics[width=\columnwidth]{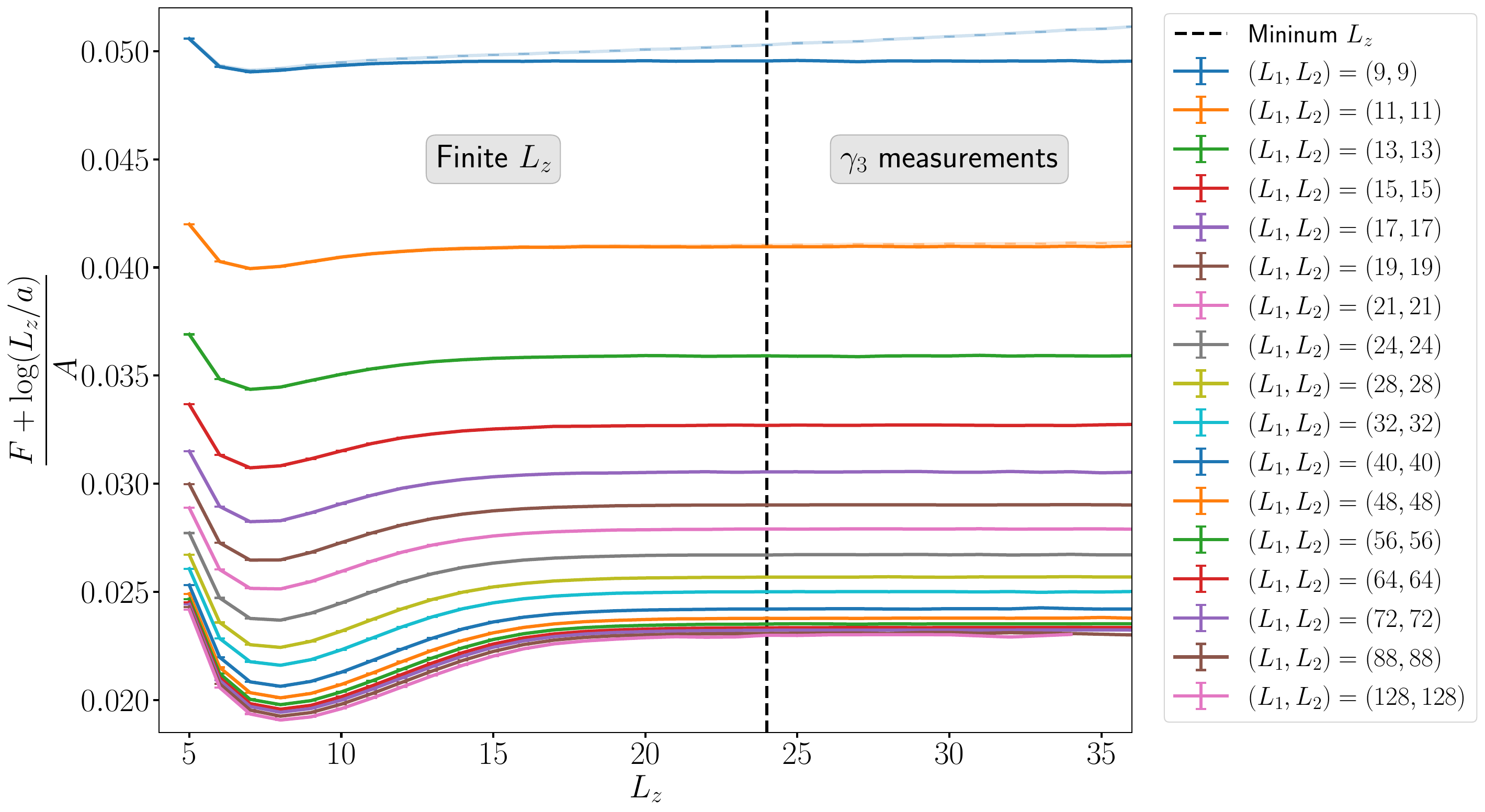}
    \caption{Comparison of the free energy estimates obtained from the dilute-gas picture
        (eq.~\eqref{eq:dilute gas}, solid lines) and the single-domain wall picture
        (eq.~\eqref{eq:single domain walls}, translucent lines). 
        The two predictions coincide except for the smallest domain wall areas, indicating that additional domain walls are strongly suppressed. 
        Simulations were performed at $\beta = 0.23$; error bars are often smaller than the line width. 
        The vertical dashed line separates the data used for studying finite transverse volume effects (left) from the data used to determine $\gamma_3$ (right). 
        The same qualitative behavior is observed at all simulated temperatures. For clarity, only half of the measured areas are shown.}
    \label{fig:dilute gas vs single domain wall}
 \vspace{-0.5cm}
\end{figure}
 
\section{Measuring the EST parameters }\label{sec:numerics}
This section outlines the strategies and intermediate steps that lead to our main result, namely the measurement
\begin{equation}
    \gamma_3(0) = -0.82(15)|\gamma_3^{\text{min}}|=-0.00106(18).
\end{equation}
where $\gamma_3^\text{min}$ is the bootstrap lower bound \cite{miroFluxTubeSmatrix2019}.
It is organised as follows.
In subsec.~\ref{subsec:dilute gas} we show that multiple domain walls are strongly suppressed and argue for the dilute gas of interfaces picture, at least for small interface areas.  
Subsec.~\ref{subsec:measuring the string tension} explains our strategy for extracting the string tension.  
The measurements themselves are presented and validated by determining the critical exponent $\nu = 0.6298(1)$, with a precision comparable to dedicated Monte Carlo studies, which reported $\nu = 0.62991(9)$ \cite{ferrenbergPushingLimitsMonte2018}.  
In subsec.~\ref{subsec:constant} we determine the area-independent contribution to the free energy—the normalisation of the partition function.  Although sometimes regarded as model-dependent, we show that for the three-dimensional Ising model it agrees with the universal prediction of effective string theory, i.e. the normalisation in eq.~\eqref{eq:partition function}.  
Subsec.~\ref{subsec:gamma3 measurements} describes how $\gamma_3$ is extracted from the numerical data and discusses possible sources of systematic uncertainty.
Finally, in subsec.\;\ref{sec:effective string tension} we measure finite-transverse-volume corrections to the string tension and argue they cannot be explained by considering the EST in finite-transverse-volume.

\subsection{A dilute gas of domain walls versus a single domain wall}\label{subsec:dilute gas}

A major source of systematic uncertainty is our limited knowledge of the microscopic dynamics—specifically, how many domain walls are present and how they interact.  This uncertainty affects the relation between the domain wall free energy and the ratio of partition functions.  Two common assumptions are employed in the literature: a single domain wall 
\begin{equation}\label{eq:single domain walls}
    F=-\ln\left(\dfrac{Z_{\text{AP}}}{Z_{\text{P}}}\right),
\end{equation}
where $a$ is the lattice spacing, or a dilute gas of non-interacting/weakly interacting domain walls \cite{hasenbuschDirectMonteCarlo1993}
\begin{equation}\label{eq:dilute gas}
    F=-\ln\left(\frac{1}{2}\ln\left(\frac{1+\nicefrac{Z_{\text{AP}}}{Z_{\text{P}}}}{1-\nicefrac{Z_{\text{AP}}}{Z_{\text{P}}}}\right)\right).
\end{equation}

Figure~\ref{fig:dilute gas vs single domain wall} shows that the two predictions agree for all but the smallest domain wall areas, $(L_x,L_y)=(8,8)$, implying that the free energy is sufficiently large to suppress additional domain walls.  The residual discrepancies at small areas likely signal the appearance of multiple domain walls. Because the dilute-gas free energy is transverse volume-independent, even for small areas with multiple domain walls, suggests they don't interact strongly with each other. Thus, we can isolate the $\mathcal{A}^{-3}$ non-universal term controlled by $\gamma_3$. Without this invariance, finite-transverse-volume effects could mask $\gamma_3$.

\subsection{The string tension}\label{subsec:measuring the string tension}
  
\begin{figure}[t]
    \includegraphics[width=\columnwidth]{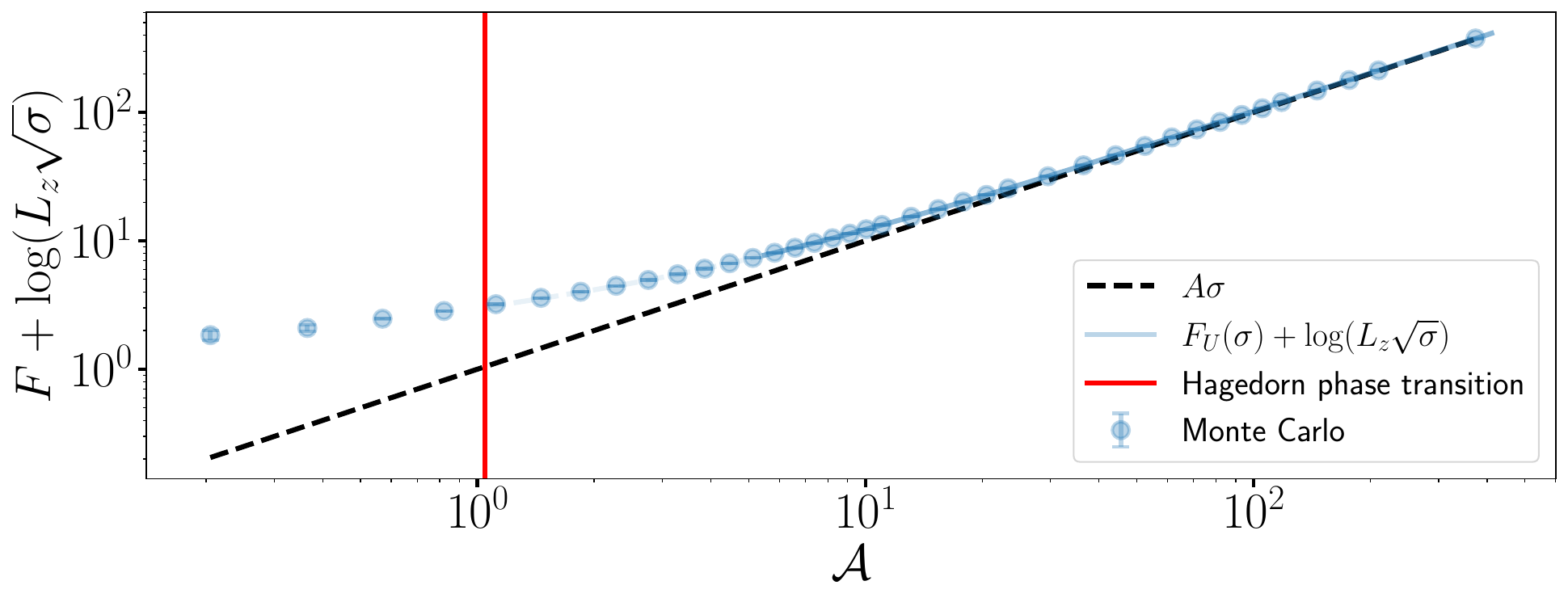}
    \caption{ Free energy versus domain wall area.
        The black dashed line shows the leading linear term; the solid blue curve is the universal contribution of the partition function.  %
        The fit only includes points with $\mathcal{A}>10$.  %
        The red vertical line marks the Hagedorn transition.}\label{fig:fit}
    \begin{minipage}[ht]{0.66\textwidth}
    \includegraphics[width=\linewidth]{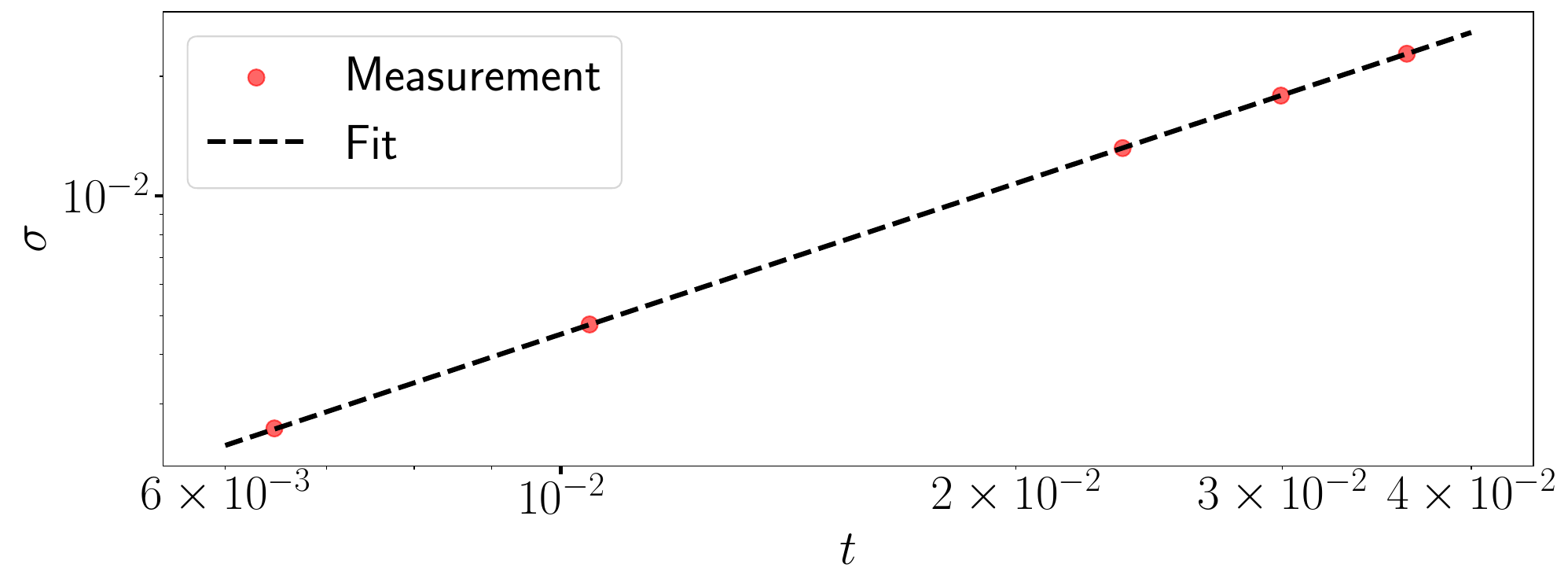}
    \caption{String tension versus reduced temperature (log-log scale).}
    \end{minipage}
    \hspace{0.1cm}
    \begin{minipage}[ht]{0.33\textwidth}
    \begin{tabular}[b]{ c|c }
    $\beta$ & $\sigma$ \\
     \hline
      0.223102 & 0.0026062(12)\\
      0.224    & 0.0047567(10)\\
      0.227    & 0.0132004(25)\\
      0.2285   & 0.0178901(14) \\
      0.23     & 0.0227944(20) 
      \vspace{1.2cm}
    \end{tabular}
     \captionof{table}{Measured string tensions. Statistical errors only.}
        \label{tab:table string tension}
   
  \end{minipage}
   \vspace{-0.5cm}\end{figure}

Determining the string tension is complicated by the unknown area at which $\gamma_3$ corrections become relevant. The free energy is fitted to eq.~\eqref{eq:partition function rectangular} with the string modes truncated at $\sim400$.\footnote{%
Even for very small areas ($\mathcal{A}\sim1.1\,\mathcal{A}_{\text{crit}}$) fewer modes would likely suffice, but retaining a large cutoff is numerically inexpensive and safer.}
To reduce statistical errors, we average the free energy, after subtracting the $\log(L_z)$ contribution, over all transverse sizes above the threshold indicated by the vertical line in fig.~\ref{fig:dilute gas vs single domain wall}.

As a first step we fit only data with sufficiently large areas (typically $\mathcal{A}>20$) to minimize putative contamination, at the cost of degraded statistics. To optimise the fit range, which we have no way of knowing \textit{a priori}, we proceed iteratively:  
(i)~compute the difference between the data and the universal prediction;  
(ii)~identify the minimum area where this difference falls below the numerical uncertainty (typically $\mathcal{A}\approx8$);  
(iii)~refit using this smaller cutoff area and verify that the deviations are smaller than the statistical uncertainty in the fitted domain.\footnote{%
Extending the fit range reduces the statistical error on~$\sigma$, which tightens the error on the difference.  We therefore increased the lower cut if necessary.}

The fit quality is illustrated in fig.~\ref{fig:fit}; the deviations are shown in fig.~\ref{fig:gamma3}.  All quoted string tensions follow this procedure (Table~\ref{tab:table string tension}).  From the temperature dependence $\sigma\sim t^{2\nu}$, excluding the first data point we obtain $\nu = 0.6298(1)$, in excellent agreement with the bootstrap result $\nu = 0.629971(4)$ \cite{kosPrecisionIslandsIsing2016} and state-of-the-art Monte Carlo simulations $\nu=0.62991(9)$ \cite{ferrenbergPushingLimitsMonte2018}. Although our precision is sufficient to resolve subleading $t$-dependent corrections to the string tension, including the critical exponents would over-parameterize the fit (fewer data points than parameters). Thus, these systematics are the leading contribution to the $\nu$ uncertainty. See \cite{caselleInterfaceFreeEnergy2007} for a detailed discussion about the subleading corrections.

\subsection{Area-independent contribution}\label{subsec:constant}

\begin{figure}[t]
    \centering
    \includegraphics[width=\columnwidth]{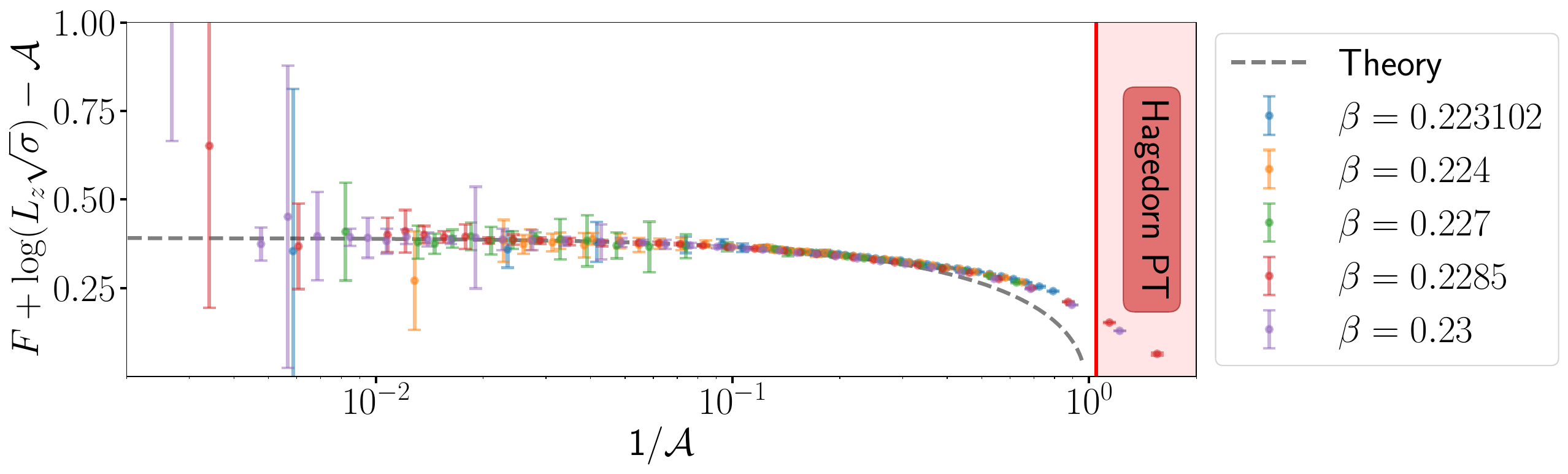}
    \caption{Measurement of $c_0\equiv\lim_{\mathcal{A}\to\infty }F(\mathcal{A})-\mathcal{A} +\log(L_z\sqrt{\sigma})$, for multiple temperatures. The dashed line is the universal contribution in eq\;\eqref{eq:c0}.
    Non universal effects close to the Hagedorn transition are  visible.}
    \label{fig:linear behaviour and const}
 \vspace{-0.5cm}\end{figure}

With the string tension $\sigma$ fixed, we define the area-independent part of the free energy 
\begin{equation}
c_0\equiv\lim_{\mathcal{A}\to\infty}\bigl[F(\mathcal{A})-\mathcal{A}+\log(L_z\sqrt{\sigma})\bigr]\label{eq:c0}
\end{equation} as illustrated in fig.~\ref{fig:linear behaviour and const}. 
For large areas the data coincide with the effective-string prediction, eq.~\eqref{eq:partition function},
\begin{equation}
    -c_0 = -\frac{1}{2}\log\left(\frac{1}{2\pi u}\right)+2\log\left|\eta \left(iu\right)\right|,
\end{equation}
after the $L_z$-dependent term is removed.
Allowing $c_0$ to float in the fit yields the same value, suggesting that the
normalisation is universal—contrary to the expectations in 
\cite{billoPartitionFunctionInterfaces2006,caselleJarzynskisTheoremLattice2016}.
Fixing $c_0$ therefore removes an unconstrained parameter and significantly sharpens the fit of the EST parameters.

\subsection{\texorpdfstring{Measuring $\gamma_3$}{ Measuring Ɣ₃}}\label{subsec:gamma3 measurements}
As explained in subsec.\;\ref{subsec:subleading TBA}, we do not use the asymptotic expansion, eq.\;\eqref{eq:Zg3 expansion}, to fit the Monte Carlo data. If one nevertheless performs that fit, one obtains $\gamma_3=-0.90(10)|\gamma_3^\text{min}|$, where $\gamma_3^\text{min}=-1/768$ is the lower  bound obtained with the S-matrix bootstrap \cite{miroFluxTubeSmatrix2019}. This result lies withing $1\sigma$ of saturating the bootstrap bound and is in clear tension with the value reported by the authors in \cite{baffigoIsingStringNambuGoto2023}. Notice that we do not expect the bound to be saturated in the case of a domain wall in the 3d Ising model, because the bound corresponds to an integrable theory, the goldstone S-matrix describing the flow from tricritical Ising to free fermions \cite{miroFluxTubeSmatrix2019}.

 \begin{figure}[tp]
    \centering
            \includegraphics[width=0.99\linewidth]{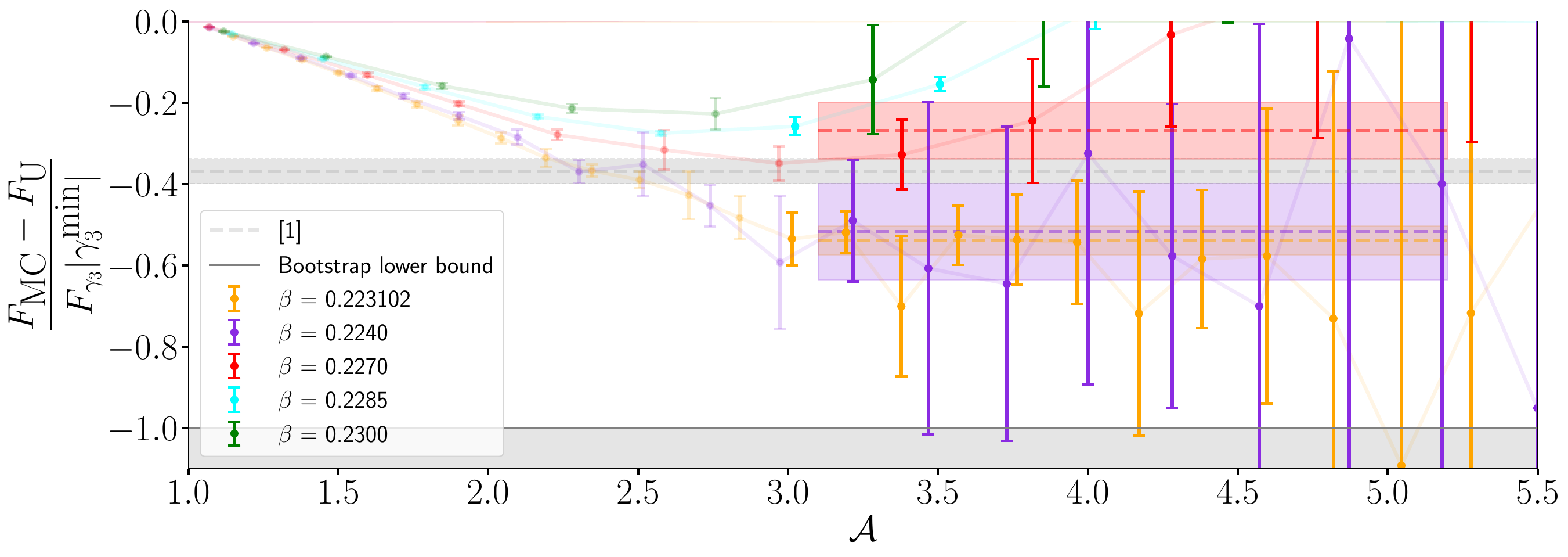}
        
        \caption{$\gamma_3^\text{MC}$ as defined in eq.\;\eqref{eq:defintion gamma_3}, normalized by the bootstrap lower bound, shown as a function of the area $\mathcal{A}$ for several inverse temperatures. Horizontal lines indicate the $\gamma_3$ value extracted from plateau fits. For $\beta=0.23$ and $\beta=0.2285$ it increases with $\mathcal{A}$ and no plateau is observed. For $\beta=0.227$ the behavior is unclear. For $\beta=0.224$ and $\beta=0.223102$, there are hints of a plateau, although it is not sharply defined due to large uncertainties. Overall, the true value will be somewhere between the orange data and the bootstrap bound.
        }
        \label{fig:gamma3}
        
        \includegraphics[width=0.99\linewidth]{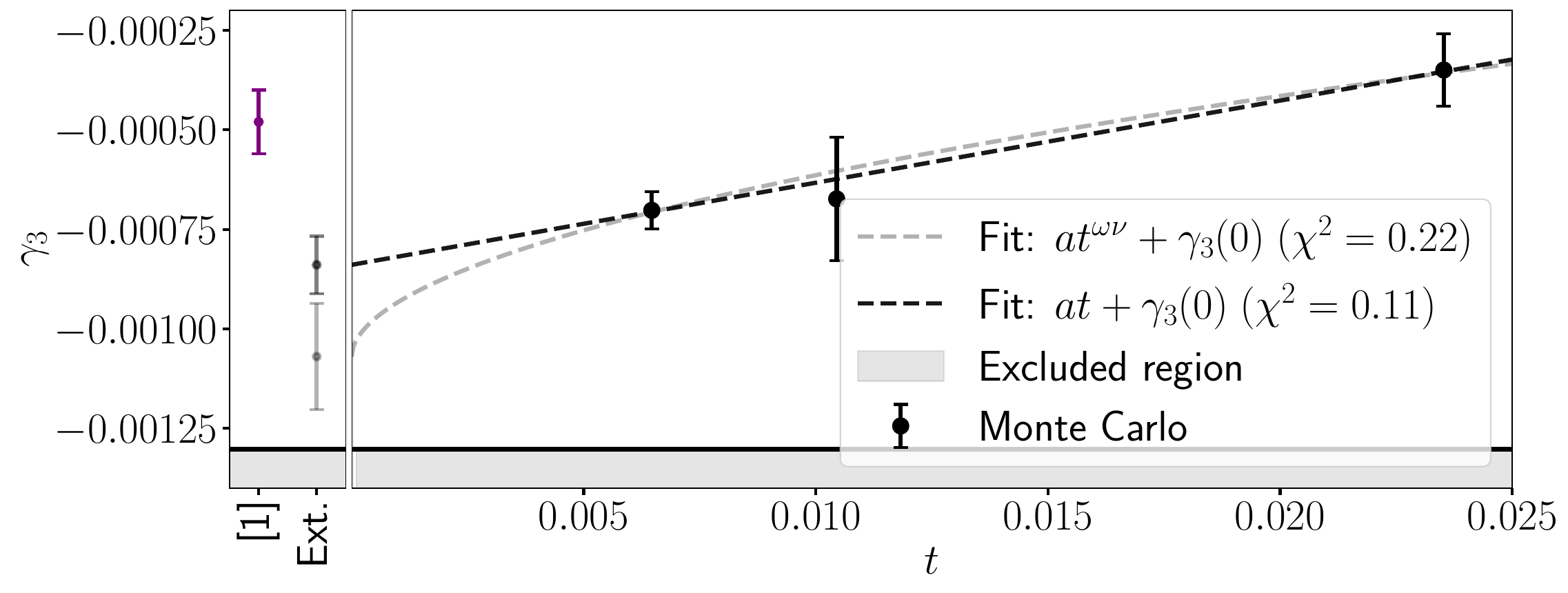}

        \caption{Extracted values of $\gamma_3$ versus reduced temperature. 
        The black line is the bootstrap lower bound \cite{miroFluxTubeSmatrix2019}.  %
        On the left subplot the ticks are:\textbf{ left)} is the state-of-the-art  \cite{baffigoIsingStringNambuGoto2023} (purple dot); \textbf{right)}: linear extrapolation.}
        \label{fig:gamma3 temperature dependence}
        
 \vspace{-0.5cm}\end{figure}

Although the large-area expansion is the natural organizing principle, the associated asymptotic series has poor convergence. We therefore expand only in \(\gamma_3\) while keeping the area dependence exact, using eq.\;\eqref{eq:definition Zg3}: 
\begin{equation}\label{eq:defintion gamma_3}
    F_\text{MC}=F_\text{U}+\gamma_3^\text{MC} F_{\gamma_3}\Longrightarrow \gamma_3^\text{MC} \equiv \dfrac{ F_\text{MC}-F_\text{U}}{F_{\gamma_3}}.
\end{equation}
Here \(F_{\text{U}}\) is obtained by numerically evaluating the first term in eq.\;\eqref{eq:full partition function} up to string level \(400\). We used the string tension measured in subsec.\;\ref{subsec:measuring the string tension}.
 \(\gamma_3^{\text{MC}}\) is shown in fig.\;\ref{fig:gamma3}.
There is only a narrow window in area where the \(\mathcal{A}^{-3}\) scaling is identifiable.

We also investigate the temperature (i.e. distance to the critical point) dependence of \(\gamma_3\). Strictly speaking, the EST description is well defined only at the critical point, where continuum symmetries are restored. However, at fixed dimensionless area $\mathcal{A}$, the size of the domain wall in lattice units diverges as we approach the continuous phase transition.
Consequently, we must determine \(\gamma_3\) at a finite distance from criticality and then extrapolate to \(\beta_c\).

 The result of the extrapolation is shown in fig.\;\ref{fig:gamma3 temperature dependence}, where $\gamma_3$ decreases monotonically as we approach the critical point. Performing a finite size scaling analysis, we expect the leading correction to be controlled by $\theta=\omega\nu$, where $\omega=0.82951(61)$ is the critical exponent of the leading irrelevant operator in the 3d Ising  \cite{Reehorst:2021hmp}. Thus, the dependence on the reduced temperature is modeled as
\begin{equation}
\gamma_3(t)=\gamma_3(0)+at^{\omega\nu},
\end{equation}
where $a$ is a fit parameter. Extrapolating to $t=0$ yields
\begin{equation}
\gamma_3(0)=-0.82(10)|\gamma_3^{\text{min}}|=-0.00106(13).
\end{equation}
However, this fit is based on only three points and therefore does not robustly test the expected 
$t$-dependence near criticality; systematic uncertainties are not controlled. As a diagnostic for systematics, we also perform a linear fit, which is indistinguishable (within errors) from the above over the available data, and obtain
\begin{equation}
\gamma_3(0)=-0.67(5)|\gamma_3^{\text{min}}|=-0.00088(6).
\end{equation}
Taking the difference between the two extrapolations as an estimate of the systematic uncertainty, we quote
\begin{equation}
\gamma_3(0)=-0.82(15)|\gamma_3^{\text{min}}|=-0.00106(18).
\end{equation}

As \(\beta\) decreases toward criticality, the \(\mathcal A\)-dependence in fig.\;\ref{fig:gamma3} evolves  from a wedge-like profile at \(\beta=0.23\) (with no sign of saturation) to a decreasing  curve that appears to level off into a plateau for \(\beta=0.224\) and \(\beta=0.223102\); the putative plateau is, however, only weakly resolved given the present uncertainties. At intermediate values (\(\beta=0.2285\)–\(0.227\)), the behavior is transitional and a plateau is not clearly identifiable.

Taken together, these trends are consistent with the emergence of a plateau only sufficiently close to the critical point—where the EST description is expected to hold most cleanly.  Further progress will likely require improved statistics in the mid-\(\mathcal A\) regime, and data closer to criticality.

\paragraph{Higher-order Wilson coefficients}\hfill \break
It is important to assess how higher-order Wilson coefficients affect the determination of $\gamma_3$. These coefficients enter the asymptotic  expansion in the area and they appear multiplied by increasingly large numerical factors. From eq.\;\ref{eq:higher order corrections.} we find that, for square domain walls, 
\begin{equation}
\dfrac{Z_{\gamma_n}}{Z_{\gamma_3}}\sim \dfrac{\Delta {\cal E}^{(n)}_{0,0,0,0}}{\Delta {\cal E}^{(3)}_{0,0,0,0}}=\dfrac{225\ 4^{2 n-3} \pi ^{2 n-6} \zeta (-n)^2}{{\cal A}^{2n-6}},
\end{equation}
which yields $\frac{25600 \pi ^4}{441 \mathcal{A}^4}$ for $n=5$. This ratio is smaller than  1  for ${\cal A}\gtrsim 100$, well beyond the range of areas where our numerical precision is sufficient. This observation  implies that, for $\gamma_3$ to remain the dominant contribution within the range of accessible areas, one must have $\gamma_5 \lesssim 10^{-3}\gamma_3$.
This order of magnitude is also suggested by the S-matrix bootstrap bounds \cite{miroFluxTubeSmatrix2019}.

\subsection{Effective string tension}
\label{sec:effective string tension}
While measuring $\gamma_3$, we determined the threshold $L_z$ above which the free energy becomes independent of $L_z$. 
Instead of discarding data points below this threshold, we argue that they offer valuable insights into the interactions between the lightest bulk particle and the domain wall, whose detailed analysis we present elsewhere.

 \begin{figure}[t]
   
    \centering
    \includegraphics[width=\linewidth]{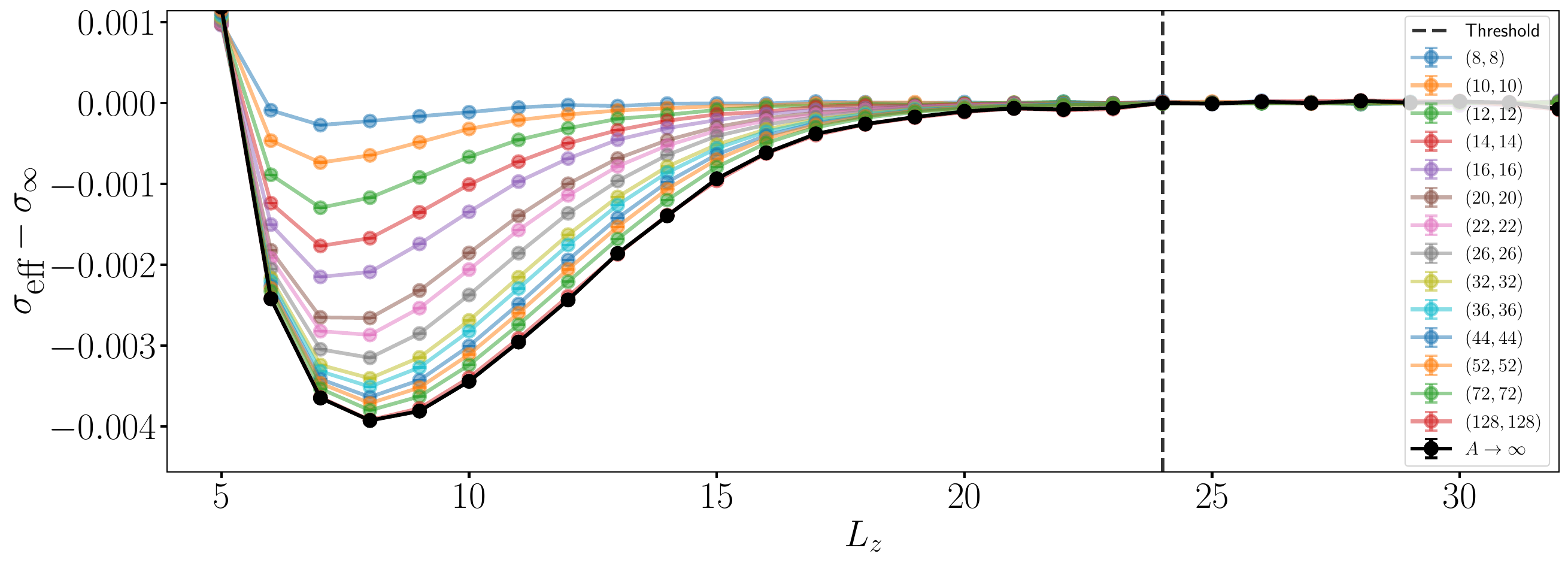}
    
    \caption{Difference between the effective string tension, defined in eq.\;\eqref{eq:effective string tension}, and the string tension as a function of $L_z$ for $\beta=0.23$.
    The vertical black dashed line marks the threshold separating the $L_z$ values used to measure $\gamma_3$.
    Colors represent measurements at finite area, while the black line shows the extrapolation to infinite area. The domain wall area increases from the top to the bottom. }
    \label{fig:effective string tension}
  \centering

    \end{figure}

Let us start by defining an effective string tension as 
\begin{equation}
    \sigma_{\text{eff}}(L_z) \equiv \lim_{A\to\infty} \dfrac{F(\mathcal{A},u,L_z;\sigma) +\log(L_z\sqrt{\sigma})}{A}.\label{eq:effective string tension}
\end{equation}
Notice that the string tension mentioned in the previous sections is $\sigma=\sigma_\infty \equiv \lim_{L_z\to\infty} \sigma_{\text{eff}}(L_z)$.  The EST predicts finite-transverse-volume corrections to the free energy studied in app.\;\ref{app:finite Lz}, which, however, vanish in the infinite-area limit, see eq.\;\ref{eq:partition function in finite-transverse-volume}. Thus
\begin{equation}
    \sigma_{\text{eff}}^\text{EST}(L_z) = \sigma_\infty,
\end{equation} a prediction inconsistent with our data, as shown in fig.\;\ref{fig:effective string tension}. We observe a trend opposite to the EST prediction: larger areas yield larger finite-transverse-volume corrections. In the curves shown, the area increases from top to bottom. We present data only for $\beta = 0.23$, which is representative of all sampled temperatures.%

\begin{figure}

\includegraphics[width=\linewidth]{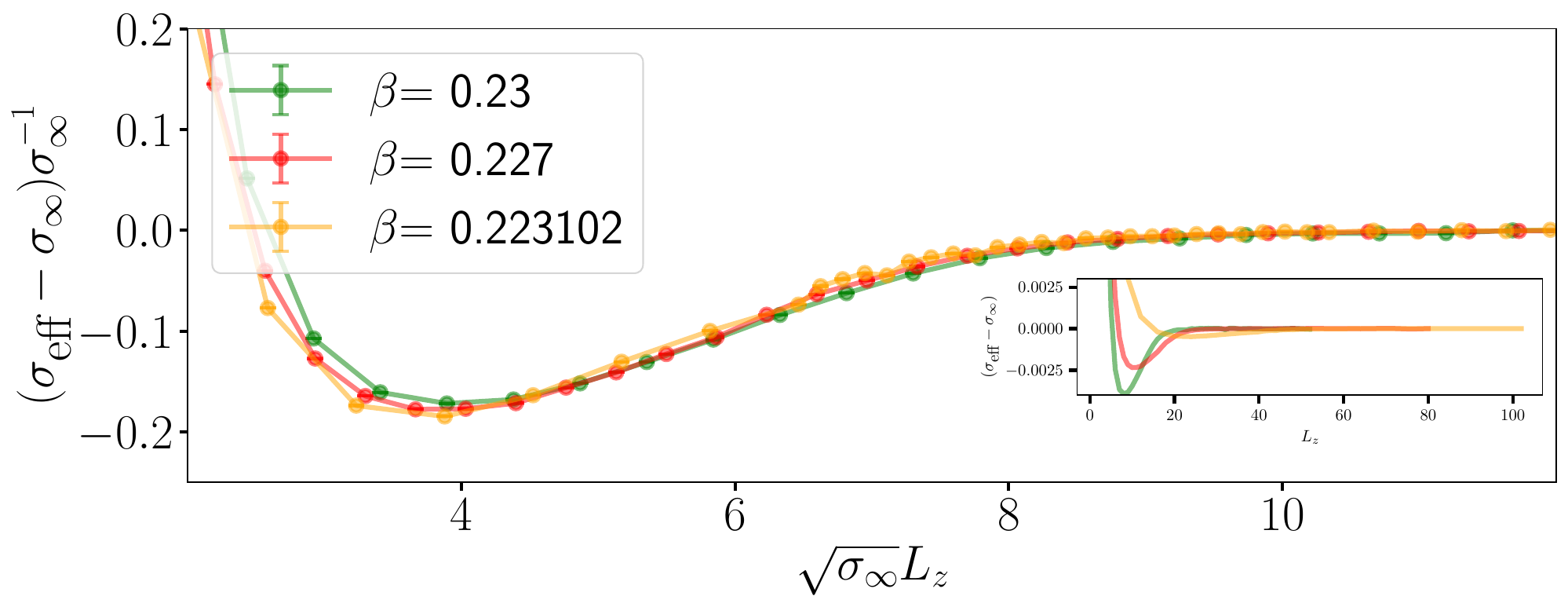}
    \caption{Rescaled string tension as a function of $\sqrt{\sigma_\infty}L_z$.
    We see that different temperatures collapse into the same curve in these rescaled variables.
    The inset  displays the original data. 
    }
    \label{fig:collapsed corrections to the effective string}
 \vspace{-0.5cm}\end{figure}

In fig.\;\ref{fig:collapsed corrections to the effective string}, we argue that these corrections are universal.
Measuring everything in units of the string tension, we see that different temperatures collapse into the same curve. In the continuum limit $\beta \to \beta_c$, this curve gives the free energy of a domain wall in the 3d Ising field theory. From the QFT perspective, this can be defined starting from the 3d Ising CFT on $\mathbb{R}^2\times S^1$ with anti-periodic ($\mathbb{Z}_2$ flipping) boundary conditions along $S^1$ and turning on the ($\mathbb{Z}_2$ even) relevant deformation into the ferromagnetic phase.

A simple feature of fig.\;\ref{fig:collapsed corrections to the effective string} is the rapid approach $\sigma_\text{eff}(L_z) \to \sigma_\infty$ when $L_z \to \infty$. Not surprisingly, this is controlled by the exponential $e^{-m L_z}$ where $m$ is the mass of the lightest bulk particle.
This observation allows us to measure the coupling between such particle and the domain wall as we shall discuss elsewhere.

\section{Conclusion}
We extended the effective string description from the long string regime to a toroidal domain wall and tested it with high-precision Monte Carlo data for the 3d Ising model. On the theory side, we clarified the normalization of the torus partition function and computed
the leading and subleading non-universal correction controlled by the Wilson coefficient $\gamma_3$, both via a worldsheet path-integral (only the leading term) and through a TBA-based derivation. %

A toroidal domain wall has two parameters: the aspect ratio $u$ and the twist $\alpha$ that control the moduli $\tau=\alpha+i u$ of the torus.
It would be interesting to explore the freedom to vary $\tau$ to explore different aspects of the EST. For example, to enhance the contribution of the leading nonuniversal terms proportional to $\gamma_3$ - see figure \ref{fig:dependence in the modular parameter}.

Numerically, we introduced an improved flat-histogram strategy that delivers precise free energy measurements over a broad range of areas and temperatures, which we validated by recovering the Ising critical exponent $\nu=0.6298(1)$, consistent with state-of-the-art Monte Carlo estimates. We also established that the area-independent contribution to the free energy matches the EST prediction, simplifying the data analysis. 

A key outcome is the determination of the leading Wilson coefficient,
\[
\gamma_3 \;=\; -0.82(15)|\gamma_3^{\text{min}}|=-0.00106(18), 
\]
obtained by expanding only in $\gamma_3$ while keeping the full area dependence. %

Beyond the EST baseline, we observed finite–transverse–volume effects that are larger than predicted by a naive finite-$L_z$ EST and exhibit a universal collapse when expressed in units of the string tension. This points to a coupling between the lightest bulk excitation and the domain wall, whose quantitative analysis we defer to future work. 

Finally, it would be interesting to quantify the first inelastic effects at ${\cal O}(\gamma_3^2/{\cal A}^{5})$. In principle, this could be achieved by comparing the free energy computed with TBA (assuming integrability) with the measured free energy using Monte Carlo. However, this requires significantly better statistics than what is possible with the current methods.

\acknowledgments

We thank  M. Caselle,  V. Gorbenko, A. Guerrieri, M. Meineri and P. Vieira for useful discussions. JM is grateful for the feedback from participants of the conference "Bridging analytical and numerical methods for quantum field theory". 
This work was supported by FCT - Fundação para a Ciência e Tecnologia, I.P. by project reference and DOI identifier \href{https://doi.org/10.54499/2021.04743.BD}{https://doi.org/10.54499/2021.04743.BD}. 
JM and JV thank the cluster time provided by INCD funded by FCT and FEDER under the grants 2021.09830.CPCA, 2023.11029.CPCA, 2024.09383.CPCA  as well as GRID FEUP. They also thank Centro de Física do Porto funded by Portuguese Foundation for Science and Technology (FCT) within the Strategic Funding UIDB/04650/2020.
JP is supported by the Simons
Foundation grant 488649 (Simons Collaboration on the Nonperturbative Bootstrap) and
the Swiss National Science Foundation through the project 200020\_197160 and through
the National Centre of Competence in Research SwissMAP. 

\appendix

\section{Computation of the Gaussian partition function}\label{sec:Gaussian partition function}
Let us consider the Gaussian partition function. 
\begin{equation}
    \tilde{Z}_0 = \int\left[D\pi\right]\exp\left(-\dfrac{1}{2}\int_{T^{2}}d^{2}\xi\left(\partial_{i}\pi\right)^{2}\right).
\end{equation}
By expanding the field in Fourier modes and imposing appropriate boundary conditions, the path integral reduces to a Gaussian integral over countably many variables. Consequently, it can be viewed as the determinant of an infinite-dimensional matrix. If the zero mode is retained, this determinant vanishes; however, the zero mode can be integrated out by hand, yielding a factor equal to the volume of the compact transverse dimension. In our case an additional factor of $\sqrt{\sigma}$ owing to the rescaling performed in subsec.\;\ref{sec:review path integral}. Thus   
\begin{equation}
    \tilde{Z}_0= \det\negmedspace^{\prime}\left[-\dfrac{\square}{2\pi}\right]^{-1/2}\sqrt{\sigma L_{z}^{2}}\,,
\end{equation}
where the prime denotes the determinant excluding the zero mode. 

The determinant can be computed using $\zeta$-function regularization \cite{faulhuberExtremalDeterminantsLaplaceBeltrami2020,dietzRenormalizationStringFunctionals1983}
\begin{equation}
    \ln\det-\dfrac{\square}{2\pi}=-\left.\dfrac{d}{ds}\left(\zeta_{-\frac{\square_{T^{2}}}{2\pi}}\left(s\right)\right)\right|_{s=0},
\end{equation}
where $\zeta_D$ is the spectral zeta function of a differential operator $D$
\begin{equation}
    \zeta_{D}\left(s\right)\equiv\sum_{n}\hspace{0cm}^{\prime}\dfrac{1}{\lambda_{n}^{s}}.
\end{equation}
Thus, for a generic torus with $\tau \equiv \alpha + iu$ 
\begin{equation}
\zeta_{-\frac{\square_{T^{2}}}{2\pi}}\left(s\right)	=\sum_{(m,n)\in\mathbb{Z}^{2}\backslash\{(0,0)\}}\dfrac{1}{\left(2\pi\dfrac{\left|m+\tau n\right|^{2}}{u}\right)^{s}}
	=2^{-s}\bar{G}_{s}\left(\tau\right).
\end{equation}
For a review of modular forms and Eisenstein series check \cite{dhokerLecturesModularForms2022}. In order to compute the derivative of the Eisenstein series at zero, it  needs to be properly analytically continued to $s=0$. For $s>1$ the series above is a real analytic Eisenstein series
\begin{equation}
    \bar{G}_{s}\left(\tau\right)\equiv\sum_{\substack{m,n\in\mathbb{Z}\\
\left(m,n\right)\neq\left(0,0\right)
}
}\dfrac{\tau_{2}^{s}}{\pi^{s}\left|m+n\tau\right|^{2s}},\label{eq:definition Eisenstein}
\end{equation}
is absolutely convergent and can be rewritten as 
\begin{equation}
    \bar{G}_{s}\left(\tau\right)	=\dfrac{1}{\Gamma\left(s\right)}\int_{0}^{\infty}\dfrac{dt}{t}t^{s}\left(\sum_{m,n\in\mathbb{Z}}e^{-\frac{\pi t}{\tau_{2}}\left|m+n\tau\right|^{2}}-1\right).
\end{equation}

In this integral representation, it is clear that the divergence of the series representation for $s<1$ arises due to the divergence of the integral near $t=0$. 
Splitting the integration domain into $[0,1]\cup[1,\infty]$, and performing the change of variables $t\to1/t$ in the first interval
\begin{equation}
    \bar{G}_{s}\left(\tau\right)=\dfrac{1}{\Gamma\left(s\right)}\left\{ \int_{1}^{\infty}\dfrac{dt}{t}t^{s}\left(\sum_{m,n\in\mathbb{Z}}e^{-\frac{\pi t}{\tau_{2}}\left|m+n\tau\right|^{2}}-1\right)+\int_{1}^{\infty}\dfrac{dt}{t}t^{-s}\left(\sum_{m,n\in\mathbb{Z}}e^{-\frac{\pi}{t\tau_{2}}\left|m+n\tau\right|^{2}}-1\right)\right\},
\end{equation}
the second term can be Poisson resumed to have the same $t$ dependence as the first
\begin{equation}
    \sum_{m,n\in\mathbb{Z}}e^{-\frac{\pi}{t\tau_{2}}\left|m+\tau n\right|^{2}}	=t\sum_{k,l\in\mathbb{Z}}e^{-\frac{t\pi}{\tau_{2}}\left|l+k\tau\right|^{2}}.
\end{equation}
Thus
\begin{equation}
    \bar{G}_{s}\left(\tau\right)\Gamma\left(s\right)=-\dfrac{1}{s\left(1-s\right)}+\int_{1}^{\infty}\dfrac{dt}{t}\left(t^{s}+t^{1-s}\right)\left(\sum_{m,n\in\mathbb{Z}}e^{-\frac{\pi t}{\tau_{2}}\left|m+n\tau\right|^{2}}-1\right),\label{eq:invariance s to 1-s}
\end{equation}
from which we conclude that $\bar{G}_{s}$ has a pole at $s=1$\footnote{The pole at $s=0$ on the right hand side is cancelled by a corresponding pole in $\Gamma(s)$.}, and  $\bar{G}_{s}\left(\tau\right)\Gamma(s)$ is invariant under $s\to1-s$. This last property is essential to compute the derivative at $s=0$
\begin{equation}
   -\left.\dfrac{d}{ds}\bar{G}_{s}\left(\tau\right)\right|_{s=0}=-\left.\dfrac{d}{ds}\dfrac{\bar{G}_{s}\left(\tau\right)\Gamma\left(s\right)}{\Gamma\left(1-s\right)}\right|_{s=1}, 
\end{equation}
where $-\left.\dfrac{d}{ds}\bar{G}_{s}\left(\tau\right)\right|_{s=1}$ is computed using the Kronecker's limit formula 
\begin{equation}
    \tilde{G}_{s}\left(\tau\right)=\dfrac{\pi}{s-1}+2\pi\left(\gamma_{E}-\log\left(2\right)-\log\left(\sqrt{\tau_{2}}\left|\eta\left(\tau\right)\right|^{2}\right)\right)+\mathcal{O}\left(s-1\right).
\end{equation}
Finally, we obtain 
\begin{equation}
\tilde{Z}_0=\left(\dfrac{i \sigma L_{z}^{2}}{\pi (\tau-\bar\tau)}\right)^{1/2}\left|\eta\left(\tau\right)\right|^{-2}\,,   
\end{equation}
which matches the leading term in \eqref{eq:partition function} (up to the trivial factor $e^{-\mathcal{A}}$).

\section{Path integral for leading non-universal free energy correction}\label{subsec:green functions}
 To simplify the regularization, we expand every index so that 
\begin{align}
    \langle V_\text{NU} \rangle = & -2\mathcal{A}\dfrac{\gamma_{3}\sigma^{2}}{\mathcal{A}^{4}}\int d^{2}\xi\left\langle \left(\partial_{i}\partial_{j}\phi\partial^{i}\partial^{j}\phi\right)^{2}\right\rangle \\
    = & - \dfrac{4\gamma_{3}\sigma^{2}}{\mathcal{A}^{3}}\left(G_{1111}(0)G_{1111}(0)+6G_{2211}(0)G_{2211}(0)+G_{2222}(0)G_{2222}(0)\right),\label{eq:Wick contractions gamma3}
\end{align}
where $G_{ijkl}(\xi^1,\xi^2)\equiv\partial_i\partial_j\partial_k\partial_l G(\xi^1,\xi^2)$, and $G$ is the propagator of a free massless boson in finite volume given in \eqref{eq:propagator}.

The series we know how to regularize are of the type
\begin{equation}
    S_1(s) = \sum_{m,n\neq\left(0,0\right)}\dfrac{m^{s}}{\frac{m^{2}}{R_{1}^{2}}+\frac{n^{2}}{R_{2}^{2}}} \qquad \qquad S_2(s) =  \sum_{m,n\neq\left(0,0\right)}\dfrac{n^{s}}{\frac{m^{2}}{R_{1}^{2}}+\frac{n^{2}}{R_{2}^{2}}},
\end{equation}
where $G_{1111}(s) = S_1(4)$ and $G_{2222}(s) = S_2(4)$. The term with crossing derivatives can be rewritten as 
\begin{align*}
G_{2211}\left(0\right) & =\lim_{s\to2}\sum_{\substack{m,n=-\infty\\
(m,n)\neq(0,0)
}
}^{+\infty}\frac{\left(2\pi im/R_{1}\right)^{s}\left(2\pi in/R_{2}\right)^{s}}{4\pi^{2}\left(\frac{m^{2}}{R_{1}^{2}}+\frac{n^{2}}{R_{2}^{2}}\right)}\\
 & =\lim_{s\to2}4\pi^{2}\left(\sum_{\substack{m,n=-\infty\\
(m,n)\neq(0,0)
}
}^{+\infty}\dfrac{1}{2}\frac{\left(\dfrac{m^{2s}}{R_{1}^{2s}}+\dfrac{2m^{s}n^{s}}{R_{1}^{s}R_{2}^{s}}+\dfrac{n^{2s}}{R_{2}^{2s}}\right)}{\frac{m^{2}}{R_{1}^{2}}+\frac{n^{2}}{R_{2}^{2}}}-\dfrac{1}{2}\sum_{\substack{m,n=-\infty\\
(m,n)\neq(0,0)
}
}^{+\infty}\dfrac{\left(\dfrac{n^{2s}}{R_{2}^{2s}}+\dfrac{m^{2s}}{R_{1}^{2s}}\right)}{\frac{m^{2}}{R_{1}^{2}}+\frac{n^{2}}{R_{2}^{2}}}\right)
\end{align*}
where the first term, when $s\to2$, is proportional to $\bar{G}_{-1}(iu)$ defined in eq.\;\eqref{eq:invariance s to 1-s}, and the second is related with $S_1(2s)$ and $S_2(2s)$. The regularized sum converges when $\Re(s)<0$.

Using the invariance under $s\to 1-s$ derived from eq.\;\eqref{eq:invariance s to 1-s}
\begin{equation}
    \bar{G}_{-s+1}(\tau) = \dfrac{\bar{G}_{s}(\tau) \Gamma(s)}{\Gamma(-s+1)},
\end{equation} 
which, since both $\Gamma(s)$ and $\bar{G}_{s}(\tau)$ are finite for $s > 1$, it implies $\bar{G}_{-s+1}(\tau)$ is zero for positive integer values of $s>1$. In particular, $\bar{G}_{-1}(\tau)=0 $. %

It is useful to define $F_s(iu)\equiv S_2(s)$. Then, it can be shown that $S_1(s)=F_{s}\left(\frac{1}{iu}\right)$. Thus, we just need to focus on computing $F_s$. We follow \cite{dietzRenormalizationStringFunctionals1983}. Since $F_s$ converges for $\Re(s)<0$, we can sum over $m$ to obtain 
\begin{equation}
    F_{s}\left(iu\right)=2\pi\zeta\left(1-s\right)+4\pi\sum_{n=1}^{\infty}\frac{n^{s-1}}{\left(e^{2\pi nu}-1\right)}.
\end{equation}
The sum converges for any $s$,  thus $F_s(iu)$ is a meromorphic function with a pole at $s=0$. The sum on the right hand side is related with the holomorphic Eisenstein series 
\begin{equation}
    \sum_{n=1}^{\infty}\dfrac{n^{s-1}}{1-q^{n}}=\dfrac{\dfrac{E_{s}\left(iu\right)}{2\zeta\left(s\right)}+1}{2}\zeta(1-s) \qquad\qquad E_s(\tau) \equiv \sum_{n,m\in\mathbb{Z},(n,m)=1} \dfrac{1}{(n + \tau m)^s}
\end{equation}
such that 
\begin{equation}
    F_{s}\left(iu\right)=-\pi\dfrac{E_{s}\left(iu\right)}{\zeta\left(s\right)}\zeta\left(1-s\right).
\end{equation}

Two useful cases are 
\begin{align*}
F_{2}\left(iu\right) & =\dfrac{1}{2\pi}E_{2}\left(iu\right)\\
F_{4}\left(iu\right) & =-\frac{3}{4\pi^{3}}E_{4}\left(iu\right).
\end{align*}

Using these identities, we obtain
\begin{equation}
    \langle V_\text{NU}\rangle = -\dfrac{32\gamma_{3} \pi^{6}}{225 \mathcal{A}^{3}}u^{4}E_{4}\left(iu\right)^2.
\end{equation}

\section{Effective string theory in finite-transverse-volume}\label{app:finite Lz}
Taking into account the winding $\omega$ around the compact dimension, the  spectrum is 
\begin{equation}
    \varepsilon_{k,k^{\prime},n,\omega}=\sqrt{1+\dfrac{4\pi}{\sigma L_{2}^{2}}\left(k+k^{\prime}-\dfrac{1}{12}\right)+\left(\dfrac{2\pi\left(k-k^{\prime}+n\omega\right)}{\sigma L_{2}^{2}}\right)^{2}+\left(\dfrac{2\pi n}{\sigma L_{2}L_{z}}\right)^{2}+\left(\dfrac{L_{z}\omega}{L_{2}}\right)^{2}}.
\end{equation}

In the large area expansion the universal partition function becomes 
\begin{equation}
    Z_\text{U} = e^{-\mathcal{A}}\sum_{k,k^{\prime},n,\omega}\sum_{\mu}\dfrac{h_{\mu}\left(k,k^{\prime},n,\omega\right)}{\mathcal{A}^{\mu}}p(k)p(k^\prime)e^{2\pi i\left(\alpha\left(k-k^{\prime}+n\omega\right)+ui\left(k+k^{\prime}-\frac{1}{12}+\frac{\pi n^{2}}{\mathcal{L}_{z}^{2}}+\frac{\omega^{2}\mathcal{L}_{z}^{2}}{4\pi}\right)\right)}.\label{eq:partition function in finite-transverse-volume}
\end{equation}
where $\mathcal{L}_z=\sqrt{\sigma}L_z$. It can be verified that $k,k^{\prime},n$, and $\omega$ appear in $h_\mu$ in specific combinations that allow the sum over them to be replaced by a differential operators acting on $Z_0$
\begin{equation}
    Z_0 \equiv  \sum_{k,k^{\prime},n,\omega}p(k)p(k^\prime)e^{2\pi i\left(\alpha\left(k-k^{\prime}+n\omega\right)+ui\left(k+k^{\prime}-\frac{1}{12}+\frac{\pi n^{2}}{\mathcal{L}_{z}^{2}}+\frac{\omega^{2}\mathcal{L}_{z}^{2}}{4\pi}\right)\right)}.
\end{equation}
Using 
\begin{align*}
2\pi \left(k+k^{\prime}-\dfrac{1}{12}\right)+\frac{2\pi^2 n^2 }{\mathcal{L}_{z}^{2}}+\dfrac{\omega^{2} \mathcal{L}_{z}^{2}}{2} & =-\partial_{u}\\
2\pi \left(k-k^{\prime}\right)+2\pi n \omega  & =-i\partial_{\alpha},
\end{align*}
the first three terms in $h_\mu$ take the following form
\begin{equation}
h_{\mu}\left(k,k^{\prime},n,\omega\right)=\left(\begin{array}{c}
1\\
2u^{2}\partial_{\tau}\partial_{\bar{\tau}}\\
2u^{3}\partial_{\tau}\partial_{\bar{\tau}}\left(u\partial_{\tau}\partial_{\bar{\tau}}+i\left(\partial_{\tau}-\partial_{\bar{\tau}}\right)\right)\\
\vdots\\
\end{array}\right)=\left(\begin{array}{c}
1\\
2u\mathcal{D}_{\left(0,0\right)}^{(1)}\left(\partial_{\tau},\partial_{\bar{\tau}}\right)\\
4u^{2}\mathcal{D}_{\left(0,0\right)}^{(2)}\left(\partial_{\tau},\partial_{\bar{\tau}}\right)\\
\vdots\\
\end{array}\right),
\end{equation}
which we explicitly checked are the modular covariant derivatives, $\mathcal{D}^{(n)}_{(0,0)}$, that take a modular form of weight $(0,0)$ into a form of weight $(n,n)$. Since $u$ has weight $(-1,-1)$ and $Z_0$ has weight $(0,0)$, we can see that each coefficient in the large area expansion is explicitly modular invariant. See \cite{aharonyModularInvarianceUniqueness2019,dhokerLecturesModularForms2022,jiangPedagogicalReviewSolvable2021a} for a review.

Thus, we obtain 
\begin{equation}
    Z_U = e^{-\mathcal{A}} \sum_{n=0} \alpha_n \dfrac{u^n}{\mathcal{A}^n } \mathcal{D}_{\left(0,0\right)}^{(n)}\left(\partial_{\tau},\partial_{\bar{\tau}}\right) Z_0,
\end{equation}
where we made the area dependence explicit, and $\alpha_n$ are numbers.
If we now compute the string tension in the large $\mathcal{A}$ limit, 
\begin{equation}
    \sigma_{\text{eff}}( L_z) \equiv \lim_{A\to\infty} \dfrac{F(A,L_z)}{A} = \sigma_\infty - \dfrac{\log \left(
    Z_0
    \right) }{A} + \mathcal{O}\left(A^{-2}\right),
\end{equation}
we observe that, in the strict $A\to\infty$, only the first term survives. 

We conclude that the EST alone cannot explain the $L_z$ dependence observed in fig.\;\ref{fig:effective string tension}.

\section{Thermodynamic Bethe Ansatz for the non-universal string}\label{app:TBA}
Our starting assumption is that, at low energies, the \(d=3\) worldsheet theory is (approximately) integrable and its \(2\!\to\!2\) branon scattering is elastic,
\[
S(s)=\exp\!\bigl[\,2i\,\delta(s)\,\bigr],\qquad
2\delta(s)=\tfrac{s}{4}+\gamma_3 s^{3}+\gamma_5 s^{5}+\cdots,
\]
where \(s\) is the (dimensionless) Mandelstam variable and we work in units with \(\sigma=1\) (hence \(\ell_s=1\)). The linear term reproduces the Nambu--Goto phase shift, while the coefficients \(\gamma_n\) are Wilson coefficients of EST. Integrability--breaking effects first appear at order \(s^{8}\) via \(2\!\to\!4\) particle production; equivalently \(|S(s)|=1-\mathcal{O}(s^{8})\), with the leading contribution proportional to \( \gamma_3^{\,2}\). The analysis below can be generalized to generic \(\{\gamma_n\}\), but here we focus on \(\gamma_3\).

To obtain finite--size energies we use \emph{excited-state TBA}: one analytically continues the ground--state TBA and inserts source terms associated with zeros crossing the integration contour (holes/strings). Solving the resulting equations yields the full finite-size spectrum for each excited state \cite{dubovskySolvingSimplestTheory2012}
\begin{align}
\varepsilon_{R}(x) & =x-i\sum_{j}2\delta\left(\frac{4x (iq_{j})}{R^{2}}\right)+\int\limits_{0}^{\infty}\frac{dx^{\prime}}{2\pi}\partial_{x^{\prime}}2\delta\left(\frac{4xx^{\prime}}{R^{2}}\right)\log\left(1-e^{-\varepsilon_{L}(x^{\prime})}\right)
\\
\varepsilon_{L}(x) & =x+i\sum_{l}2\delta\left(\frac{4x (-i p_{l})}{R^{2}}\right)+\int\limits_{0}^{\infty}\frac{dx^{\prime}}{2\pi}\partial_{x^{\prime}}2\delta\left(\frac{4xx^{\prime}}{R^{2}}\right)\log\left(1-e^{-\varepsilon_{R}(x^{\prime})}\right)
\\
E&=R+\dfrac{\sum_{l}p_{l}}{R}+\dfrac{\sum_{j}q_{j}}{R}+\dfrac{1}{2\pi R}\int_{0}^{\infty}dx\log\left(1-e^{-\varepsilon_{R}\left(x\right)}\right)+\dfrac{1}{2\pi R}\int_{0}^{\infty}dx\log\left(1-e^{-\varepsilon_{L}\left(x\right)}\right)\label{eq:TBA energy}
\\
P&=\dfrac{\sum_{l}p_{l}}{R}-\dfrac{\sum_{j}q_{j}}{R}+\dfrac{1}{2\pi R}\int_{0}^{\infty}dx\log\left(1-e^{-\varepsilon_{R}\left(x\right)}\right)-\dfrac{1}{2\pi R}\int_{0}^{\infty}dx\log\left(1-e^{-\varepsilon_{L}\left(x\right)}\right),
\end{align}
where $p_l,\; n_l$ and $\varepsilon_R$ are the momenta, monodromies and pseudo-energies of the right-movers, and $q_j\; m_j$ and $\varepsilon_L$  the analogous for the left-movers. The second and forth equation, without the sum, are the ground state TBA equations while the first and third equation, as well as the sums,  arise in the excited-state TBA from contour deformations: when the integration contour crosses zeros of $1-e^{-\varepsilon_{R,L}}$, the integrals pick up source terms located at
\[
\varepsilon_R(ip_l)=2\pi i \,n_l,\qquad 
\varepsilon_L(iq_j)=-2\pi i \,m_j,\qquad n_l,m_j\in\mathbb{Z}.
\]

To illustrate the procedure, we begin by re-deriving the GGRT spectrum (i.e.\ $\gamma_3=0$), following \cite{dubovskyFluxTubeSpectra2015c}. This serves as a benchmark for the subsequent $\gamma_3\neq 0$ analysis.

\subsection{The GGRT spectrum}
For the GGRT case the phase shift is $2\delta(s)=s/4$. In units $\sigma=1$, let $R=L_2$ so that $R^2=\mathcal{A}/u$. The system reduces to
\begin{align}
\varepsilon_{R}(x) & = x\!\left(1+\frac{P^{(1)}_L}{R^{2}}+\frac{1}{R^{2}}\int_{0}^{\infty}\frac{dx'}{2\pi}\,\log\!\left(1-e^{-\varepsilon_{L}(x')}\right)\right),\\
\varepsilon_{L}(x) & = x\!\left(1+\frac{P^{(1)}_R}{R^{2}}+\frac{1}{R^{2}}\int_{0}^{\infty}\frac{dx'}{2\pi}\,\log\!\left(1-e^{-\varepsilon_{R}(x')}\right)\right),
\end{align}
where $P^{(1)}_R\equiv\sum_l p_l$ and $P^{(1)}_L\equiv\sum_j q_j$.

The $x$-dependence of the pseudo-energies factorizes, so we set
\begin{equation}
\varepsilon_{R}(x)=c_{R}\,x,\qquad \varepsilon_{L}(x)=c_{L}\,x.
\end{equation}
Using
\[
\int_{0}^{\infty}\frac{dx}{2\pi}\,\log\!\left(1-e^{-c\,x}\right) = -\,\frac{\pi}{12\,c},
\]
the system becomes
\begin{align}
2\pi n_{l} & = p_{l}\,c_{R},\\
c_{R} & = 1+\frac{P^{(1)}_L}{R^{2}}-\frac{\pi}{12\,c_{L}\,R^{2}},\\
2\pi m_{j} & = q_{j}\,c_{L},\\
c_{L} & = 1+\frac{P^{(1)}_R}{R^{2}}-\frac{\pi}{12\,c_{R}\,R^{2}}.
\end{align}
From the first and third equations,
\begin{equation}
P^{(1)}_R=\frac{2\pi}{c_{R}}\,N_{R},\qquad
P^{(1)}_L=\frac{2\pi}{c_{L}}\,N_{L},
\end{equation}
with $N_R\equiv\sum_l n_l$ and $N_L\equiv\sum_j m_j$. Solving for $c_L$ and $c_R$ and inserting in the TBA energy,
\begin{align}
    E_{k,k'}&=R\,(c_{L}+c_{R}-1)=R\,\mathcal{E}_{k,k'}\\
    P_{k,k'}&=R(c_R-c_L)
\end{align}
one obtains
\begin{align}
\mathcal{E}_{k,k'} &= \sqrt{\,1
+\frac{4\pi}{R^{2}}\!\left(k+k'-\frac{1}{12}\right)
+\frac{4\pi^{2}}{R^{4}}\!\left(k-k'\right)^{2}\,}\,\\
P_{k,k^\prime} &=\frac{2 \pi  (k-k^\prime)}{R},
\end{align}
where $k,k'\in\mathbb{N}_0$ and the degeneracies are the partition numbers $p(k)$ and $p(k')$ (for a single transverse boson in $d=3$).

\subsection{The non-universal spectrum}
For $\gamma_3\neq0$ the logic is the same, so we only highlight the differences.
The phase shift becomes $2\delta(s)=\tfrac{s}{4}+\gamma_3 s^3$, and the pseudo-energies acquire cubic terms:
\begin{align}
    \varepsilon_{R}(x)=&x\left(1+\dfrac{1}{R^{2}}P^{(1)}_L+\dfrac{1}{R^{2}}\int\limits_{0}^{\infty}\frac{dx^{\prime}}{2\pi}\log\left(1-e^{-\varepsilon_{L}(x^{\prime})}\right)\right)\\    &+x^{3}\gamma_{3}\left(64\dfrac{P_{L}^{\left(3\right)}}{R^{6}}+\dfrac{96}{\pi R^{6}}\int\limits_{0}^{\infty}dx^{\prime}x^{\prime2}\log\left(1-e^{-\varepsilon_{L}(x^{\prime})}\right)\right),
\end{align}
where $P^{(3)}_L\equiv \sum_j q_j^3$. Working to linear order in $\gamma_3$,
\begin{align*}
\varepsilon_{R}(x) & =xc_{R}^{1,0}+x\gamma_{3}c_{R}^{1,1}+\gamma_{3}x^{3}c_{R}^{3,0}, & \varepsilon_{L}(x) & =xc_{L}^{1,0}+x\gamma_{3}c_{L}^{1,1}+\gamma_{3}x^{3}c_{L}^{3,0}\\
p_{l} & =p_{l}^{\left(0\right)}+\gamma_{3}p_{l}^{\left(1\right)},&P^{(1)}_R & =\sum_l p_l^{(0)} +\gamma_{3}\sum p_{l}^{\left(1\right)}=P_{R}^{1,0}+\gamma_3 P_R^{1,1}\\
P_{R}^{\left(3\right)} & =\sum_{l}\left(p_{l}^{\left(0\right)}\right)^{3}+3\gamma_{3}\sum_{l}\left(p_{l}^{\left(0\right)}\right)^{2}p_{l}^{\left(1\right)},
\end{align*}
where the $\gamma_3$ term in $P^{(3)}_L$ can be dropped at this order as it always appears multiplied by $\gamma_3$. Collecting terms yields the $\gamma_3=0$ set of equations
\begin{align*}
2\pi N_R=c_R^{1,0} P_R^{1,0},\quad
c_R^{1,0}=1+\frac{P_L^{1,0}}{R^2}-\frac{\pi}{12 c_L^{1,0}R^2},\quad
2\pi N_L=c_L^{1,0} P_L^{1,0},\quad
c_L^{1,0}=1+\frac{P_R^{1,0}}{R^2}-\frac{\pi}{12 c_R^{1,0}R^2},
\end{align*}
and the linear-in-$\gamma_3$ set
\begin{align*}
-c_{R}^{1,0}P_{R}^{1,1} & =c_{R}^{1,1}P_{R}^{1,0}-P_{R}^{\left(3,0\right)}c_{R}^{3,0}\\
c_{R}^{1,1} & =\dfrac{P_{L}^{1,1}}{R^{2}}+\frac{\pi^{3}c_{L}^{3,0}}{30\left(c_{L}^{1,0}\right)^{4}R^{2}}+\frac{\pi c_{L}^{1,1}}{12\left(c_{L}^{1,0}\right)^{2}R^{2}}\\
c_{R}^{3,0} & =-64\dfrac{P_{L}^{\left(3\right)}}{R^{6}}-\frac{32\pi^{3}}{15\left(c_{L}^{1,0}\right)^{3}R^{6}}\\
\vdots 
\end{align*}
where the omitted equations are obtained by swapping $R$ with $L$.

\begin{table}[t]
    \centering
    \begin{tabular}{c|c|c}
        $k$ & $s$ & $p\left(k,s\right)$ \\\hline
        0 & 0 & 1\\
        1 & 1 & 1\\
        2 & $\left(2,8\right)$ & $\left(1,1\right)$\\
        3 & $\left(3,9,27\right)$ & $\left(1,1,1\right)$\\
        4 & $\left(4,10,16,28,64\right)$ & $\left(1,1,1,1,1\right)$\\
        5 & $\left(5,11,17,29,35,65,125\right)$ & $\left(1,1,1,1,1,1\right)$\\
        \vdots & \vdots & \vdots\\
        8 & $\left(8,14,20,26,\textbf{32},38,44,56,62,68,\cdots,512\right)$ & $\left(1,1,1,1,\textbf{2},1,1,1,1,1,\cdots,1\right)$,
        \end{tabular}
    \caption{Allowed values of $s$ for each $k$ and the corresponding degeneracies. The bold values highlight the first degenerate state $8=2+2+2+2=3+1+1+1+1+1.$}\
    \label{tab:degeneracies}
\end{table}

The energy, eq.\;\eqref{eq:TBA energy}, can be rewritten as 
\begin{align}
E_{k,k^\prime,s,s^\prime}&=R\left(c_{R}^{1,0}+c_{L}^{1,0}+\gamma_{3}c_{R}^{1,1}+\gamma_{3}c_{L}^{1,1}-1\right)\\
 &=R\left(\mathcal{E}_{k,k^{\prime}}-\frac{32\pi^{6}\gamma_{3}}{225R^{8}}\frac{\left(240s+1\right)\left(240s^{\prime}+1\right)}{\mathcal{E}_{k,k^{\prime}}\left(\left(\mathcal{E}_{k,k^{\prime}}+1\right)^{2}-\dfrac{\pi^{2}}{R^{4}}\left(k-k^{\prime}\right)^{2}\right)^{3}}\right),
\end{align}
with $s\equiv\sum_l n_l^3$ and $k\equiv\sum_l n_l$. The degeneracy $p(k,s)$ is the number of partitions of $k$ whose sum of cubes is $s$, with generating function
\begin{equation}\label{eq:generating function}
    \mathcal{G}\left(x,z\right)=\prod_{m=1}\dfrac{1}{1-x^{m}z^{m^{3}}}=\sum_{k,s}p\left(k,s\right)x^{k}z^{s}.
\end{equation}
For reference, the first allowed values of $s$ for each $k$, as well as its degeneracies are in tab.\;\ref{tab:degeneracies}. As highlighted in bold, knowing all values of $s$ is not enough, since the degeneracies are non-trivial.  

The non-perturbative prediction for the ground-state energy at finite $R$ is 
\begin{equation}
    E_0(R)=R\left(\sqrt{1-\dfrac{\pi}{3R^{2}}}-\frac{32\pi^{6}\gamma_{3}}{225R^{8}}\frac{1}{\sqrt{1-\dfrac{\pi}{3R^{2}}}\left(\sqrt{1-\dfrac{\pi}{3R^{2}}}+1\right)^{6}}\right),
\end{equation}
whose large-$R$ expansion agrees with \cite{miroFluxTubeSmatrix2019, aharonyCorrectionsNambuGotoEnergy2010}.

\section{Computing the  $\gamma_3$ corrections to the partition function}\label{app:z0gamma3}
Our goal is to evaluate
\begin{equation}\label{eq:fullz0}
Z_{3}^{0}\left(\tau\right)=\left|q^{-1/24}\sum_{k,s}(240s+1)p\left(k,s\right)q^{k}\right|^{2},
\end{equation}
where $p(k,s)$ counts partitions of $k$ whose sum of cubes equals $s$. The inner sum splits as
\begin{equation}\label{eq:simplified z0}
q^{-1/24}\sum_{k,s}\left(240s+1\right)p\left(k,s\right)q^{k}=q^{-1/24}240\sum_{k,s}sp\left(k,s\right)q^{k}+\eta\left(\tau\right)^{-1},
\end{equation}
Thus, it suffices to compute $\mathcal{U}(q)=\sum_{k,s} s\,p(k,s)\,q^{k}$.
Fortunately, this is related to the generating function $\mathcal{G}$ in eq.\;\eqref{eq:generating function}, %
\begin{equation}
    \mathcal{U}\left(q\right)\equiv\left.\dfrac{d}{dz}\mathcal{G}\left(q,z\right)\right|_{z=1}=\sum_{k}q^{k}\sum_{s}sp\left(k,s\right)=\left(\prod_{m=1}^{\infty}\dfrac{1}{1-   q^{m}}\right)\left(\sum_{k=1}\dfrac{k^{3}q^{k}}{1-q^{k}}\right).
\end{equation}
The first term on the right hand side is the generating function for the partition of $k$ 
\begin{equation}
    \prod_{m=1}^{\infty}\dfrac{1}{1-q^{m}}=\sum_{m=0}^{\infty}p(k)q^{k} =\dfrac{q^{1/24}}{\eta\left(\tau\right)},
\end{equation}
with $q=e^{2\pi i \tau}$, and the second term is identified as
\begin{equation}
     \sum_{k=1}\dfrac{k^{3}q^{k}}{1-q^{k}} = \dfrac{2 (E_4(q) - 1)}{\zeta(-3)}.
\end{equation}

Plugging this back into eq.\;\eqref{eq:fullz0}, we get
\begin{equation}
    Z_{\gamma_3}^{0}(\tau)=\left|\frac{E_{4}(\tau)}{\eta(\tau)}\right|^{2}.
\end{equation}
All manipulations are valid as formal power series; analytically they converge for $|q|<1$.

\section{On zeta-function regularization}\label{app: zeta function regularization}

As shown in ref.~\cite{dubovskyEffectiveStringTheory2012}, $\zeta$-function regularization (ZFR) requires adding a non-covariant counterterm at one loop. In our approach this is not an issue; we assume from the outset that the world-sheet theory is integrable, with the GGRT spectrum, rather than a free boson deformed by an infinite tower of operators. Had we computed the universal contribution perturbatively in the large area expansion, the Polchinski–Strominger operator required to restore Lorentz invariance would have produced an additional contribution at order $\mathcal{A}^{-3}$, shifting the value of $\gamma_3$.

Ref.~\cite{aharonyModularInvarianceUniqueness2019} further argues that, under mild technical assumptions, the $T\bar{T}$ deformation is the \emph{unique} modular-invariant deformation of a two-dimensional CFT that depends on a single parameter—the string tension—and on the energy–momentum of the undeformed theory. Consequently, the universal contribution we computed cannot be further renormalized by local counterterms. Operationally, all universal contributions are evaluated (numerically) by explicitly summing over the string energy levels when fitting the numerical results.

We utilize ZFR solely to compute the leading \emph{non-universal} term. Higher-order non-universal contributions may require an explicit regulator; however, all terms at order $\mathcal{A}^{-3}$ should be already properly regulated in this mixed approach. The first potential issue arises when evaluating the cross term involving the PS operator (or higher-order counterterms) and the leading $\gamma_{3}$ correction to the action; such terms enter at order $\mathcal{A}^{-5}$ and beyond.

The failure of ZFR can be demonstrated explicitly by evaluating expectation values of the equations of motion (neglecting contact terms):
\begin{align}
   \bigl\langle \phi \square \phi \bigr\rangle &=E_0(\tau)= -1,\\
   \bigl\langle \square \phi \, \square \phi \bigr\rangle &= 0,\\
   \bigl\langle (\square^{p} \phi)^{n} \bigr\rangle &= 0, \quad p>1.
\end{align}
The pathology appears when only two derivatives act on a propagator—for instance, in $\langle \phi^{2n+1} \square \phi \rangle$ or in the PS operator. In our $\gamma_{3}$ calculation this problem does not arise, as every propagator carries four derivatives.

\bibliographystyle{JHEP}
\bibliography{All.bib,ExtraBib}

\end{document}